\documentclass{cmspaper}
\usepackage{graphicx,epsfig,rotating}
\begin{document}
\newcommand{\nc}{\newcommand}
\nc{\lsim}{\mbox{\raisebox{-.6ex}{~$\stackrel{<}{\sim}$~}}}
\nc{\gsim}{\mbox{\raisebox{-.6ex}{~$\stackrel{>}{\sim}$~}}}
\nc{\esim}{\mbox{\raisebox{-.6ex}{~$\stackrel{-}{\sim}$~}}}
\nc{\beq}{\begin{equation}}
\nc{\eeq}{\end{equation}}

\newcommand{\ra}{\rightarrow}

  \begin{titlepage}

   \cmsnote{2001/032}
   \date{15 November 2001}

\title{ Summary of the CMS Discovery Potential for the MSSM SUSY Higgses}

  \begin{Authlist}
     D. ~Denegri\Aref{a}, ~V. ~Drollinger\Aref{b}, ~R.~Kinnunen\Aref{c},
 ~K.~Lassila-Perini\Aref{c}, ~S. ~Lehti\Aref{c}, ~F. ~Moortgat\Aref{d},
 ~A. ~Nikitenko\Aref{e}, ~S. ~Slabospitsky\Aref{f}, ~N. ~Stepanov\Aref{g}
  \end{Authlist}

\Anotfoot{a} {European Laboratory for Particle Physics (CERN), 
         Switzerland and DAPNIA, Saclay, France}
\Anotfoot{b}{IEKP, Karlsruhe University, Germany - 
Now at Department of Physics and Astronomy, 
University of New Mexico, USA}
\Anotfoot{c} {Helsinki Institute of Physics, Helsinki, Finland}
\Anotfoot{d} {University of Antwerpen, Wilrijk, Belgium - Research Assistant
of the FWO, Belgium}
\Anotfoot{e}{Imperial College, University of London, London, UK - On leave from ITEP, Moscow, Russia}
\Anotfoot{f}{IHEP, Protvino, Moscow Region, Russia}
\Anotfoot{g}{European Laboratory for Particle Physics (CERN) - On leave from ITEP, Moscow, Russia}



  \begin{abstract}

This work summarises the present understanding of the expected MSSM SUSY
 Higgs reach for CMS. Many of the studies presented here result from 
detailed detector simulations incorporating final CMS detector design
and response. With $30~fb^{-1}$ the $h \ra \gamma\gamma$ and 
$h\ra b\overline{b}$ channels allow to cover most of the MSSM parameter 
space. For the massive $A,H,H^{\pm}$ ~MSSM Higgs states the channels $A,H 
\ra \tau\tau$ and $H^{\pm} \ra \tau\nu$ turn  out to be the most profitable 
ones in terms of mass reach and parameter space coverage. Consequently CMS
has made a big effort to trigger efficiently on $\tau$'s. Provided neutralinos and sleptons are not too heavy, there is an interesting complementarity
in the reaches for $A,H \ra \tau\tau$ and $A,H \ra \chi\chi$.
  \end{abstract}

  
  \end{titlepage}


  \section{Introduction}

For several years the LEP and the Tevatron have been functioning in an energy range allowing the direct search for the Higgs boson(s). LEP has recently been closed without solid evidence for the Higgs boson. The measurements
 yield as lower bounds
 114.1~GeV for the Standard Model (SM) Higgs and 91.0 and 91.9~GeV for the
 light (h) and the pseudoscalar (A) Higgs 
bosons of the MSSM Model \cite{LEPII}. The excluded $tan\beta$ regions are 
 0.5$< tan\beta <$ 2.4 for the maximal $m_h$ scenario
and 0.7$< tan\beta <$ 10.5 for the no mixing scenario \cite{LEPII,junk}. 
The indirect searches, performing fits to all existing electroweak data indicate a light SM Higgs boson with the most probable mass $m_H$ = 88~GeV+53~GeV-21~GeV
 and a 95\%
CL upper limit of 196~GeV \cite{LEPE}. Until the arrival of the LHC the only 
place where the Higgs search can be pursued in the coming yeares is at the 
Fermilab Tevatron. Integrated luminosities of from 20 to 40~$fb^{-1}$ are 
needed to explore the $m_H \lsim$ 170~GeV range. This could possibly be 
achieved by 2007-2008 but evidence for the Higgs at the Tevatron in either
the $H \ra b\overline{b}$ or $H \ra WW \ra \ell\ell\nu\nu$ channels will be marginal.
 At LHC a clear signal for the Higgs boson can be 
expected already after few months of running ($\sim$ 10~$fb^{-1}$)
 in a part of the parameter space 
($m_H \gsim 2m_Z$ for SM Higgs and high $tan\beta$ for MSSM Higgses). It is the region $m_H \lsim$ 125~GeV relying on $H \ra \gamma\gamma$ and
 $H \ra b\overline{b}$ channels which is the hardest, requiring 
$\gsim 30~fb^{-1}$. For the SM Higgs in the mass range from 125 to 180~GeV
the channels $H \ra ZZ^* \ra 4\ell^{\pm}$ \cite{puljak} 
and $H \ra WW \ra \ell\ell\nu\nu$ \cite{dittmar}
are the most appropriate ones and require from 10 to 30~$fb^{-1}$. 

In this note we present the summary of the expected reach for the following discovery channels of the MSSM Higgs bosons: 
  
$\bullet$ $h \ra \gamma\gamma$, inclusive production and production in association with an isolated
lepton in $Wh$ and $t\overline{t}h$ final states

$\bullet$ $h \ra b\overline{b}$ in association with an isolated
lepton and b-jets in $Wh$ and $t\overline{t}h$  

$\bullet$ $A, H \ra \mu\mu$, inclusive and in $b\overline{b}H/A$ final states 

$\bullet$ $A, H \ra \tau\tau$ with $2~lepton$, $lepton+\tau~jet$ 
and $2~\tau~jet$ final states

$\bullet$ $H^{\pm} \ra \tau\nu$ in $gb \ra tH^{\pm}$ and in $q\overline{q}' \ra H^{\pm}$ 

$\bullet$ $H^{\pm} \ra tb$ in $gb \ra tH^{\pm}$ 

$\bullet$ $A, H \ra \chi^{0}_{2}\chi^{0}_{2}  \ra 4\ell^{\pm}+X$     
 
Several other channels have been investigated in CMS. Some of them, 
like $qq \ra qqh$ with $h \ra \tau\tau$, $h \ra \gamma\gamma$ or $h \ra
WW$, are potentially very interesting and presently under study,
 but no conclusive results have been obtained
yet. For other channels like $h, H \ra ZZ^* \ra 4\ell^{\pm}$, the discovery ranges are significantly less important or limited to the low $tan\beta$ region being
thus either already
excluded by the LEPII results or made less plausible. 

The results presented in this note are not final ones but rather a snapshot
of our present understunding, new developments are to be expected as the full
 simulation study with present and final CMS detector design and performance
 is going on or has to be
started in the near future on most of the above channels.  

\vspace{ 3mm}
\section{Simulation tools}

The PYTHIA versions 5.7 \cite{pythia57} and 6.1 \cite{pythia} are used to 
generate physics events for the signals and for the backgrounds. With the PYTHIA version 5.7 the SUSY Higgs masses and couplings are corrected using calculations including two-loop/RGE improved radiative corrections \cite{gunion}. Polarized $\tau$ decays are incorporated using the TAUOLA package \cite{tauola}. In some
cases the HDECAY package \cite{hdecay} is used to calculate or normalize the cross sections
and branching ratios. Detector simulation is performed in general with 
the fast simulation
package CMSJET \cite{cmsjet}. However, the detector dependent issues 
like the resolutions of the narrow mass
states, missing transverse energy resolution, b - and $\tau$-tagging and jet resolution
are studied with full GEANT-based (CMSIM and ORCA) detector simulation. The 
results are either used directly or parametrized for the fast simulation study.

\vspace{ 3mm}
\section{MSSM Higgs bosons}

 With the assumption that the whole sparticle spectrum is heavy enough,
 the spectrum of the heavy MSSM Higgs bosons, the CP-odd ($A$), 
the CP-even ($H$) and the two charged ($H^{\pm}$) Higgs bosons,
 can be expressed with only two parameters, $m_A$ and $tan\beta$. In the
so-called 
decoupling limit, $m_A >> m_Z$, the heavy Higgs bosons become degenerate in 
mass. At tree-level the mass of 
the lightest CP-even Higgs boson $h$ is bound to be lighter than the $Z$,
 but the loop corrections 
to $m_h$ are large and particularly sensitive to the mass of the top quark, 
SUSY scale and 
the amount of stop-quark mixing. With the recent calculations including two-loop radiative 
corrections the upper bound of $m_h$ is $\sim$ 113~GeV with no stop mixing and if the SUSY 
scale is taken to be 1 TeV, whilst for maximal stop mixing the upper bound is $\sim$ 130~GeV.
 In the majority of our studies we assume the SUSY parameter spectrum used for the LEP
benchmark scenarios \cite{LEP}. For the no-mixing scenario the SUSY parameters are as
follows: $M_1$ = 100~GeV, $M_2$ = 200~GeV, $\mu$ = -200~GeV, $M_{\tilde{g}}$ = 800~GeV, 
$M_{\tilde{q},{\tilde{\ell}}}$ = 1 TeV and $A_t$ = 0. For the maximal $m_h$ (maximal mixing)
 $A_t$ = $\sqrt{6}\times M_{\tilde{q},\tilde{g},\tilde{\ell}}$ 
while the other parameters remain the same.

\begin{figure}[t]
\begin{center}
\resizebox{130mm}{100mm}{\includegraphics{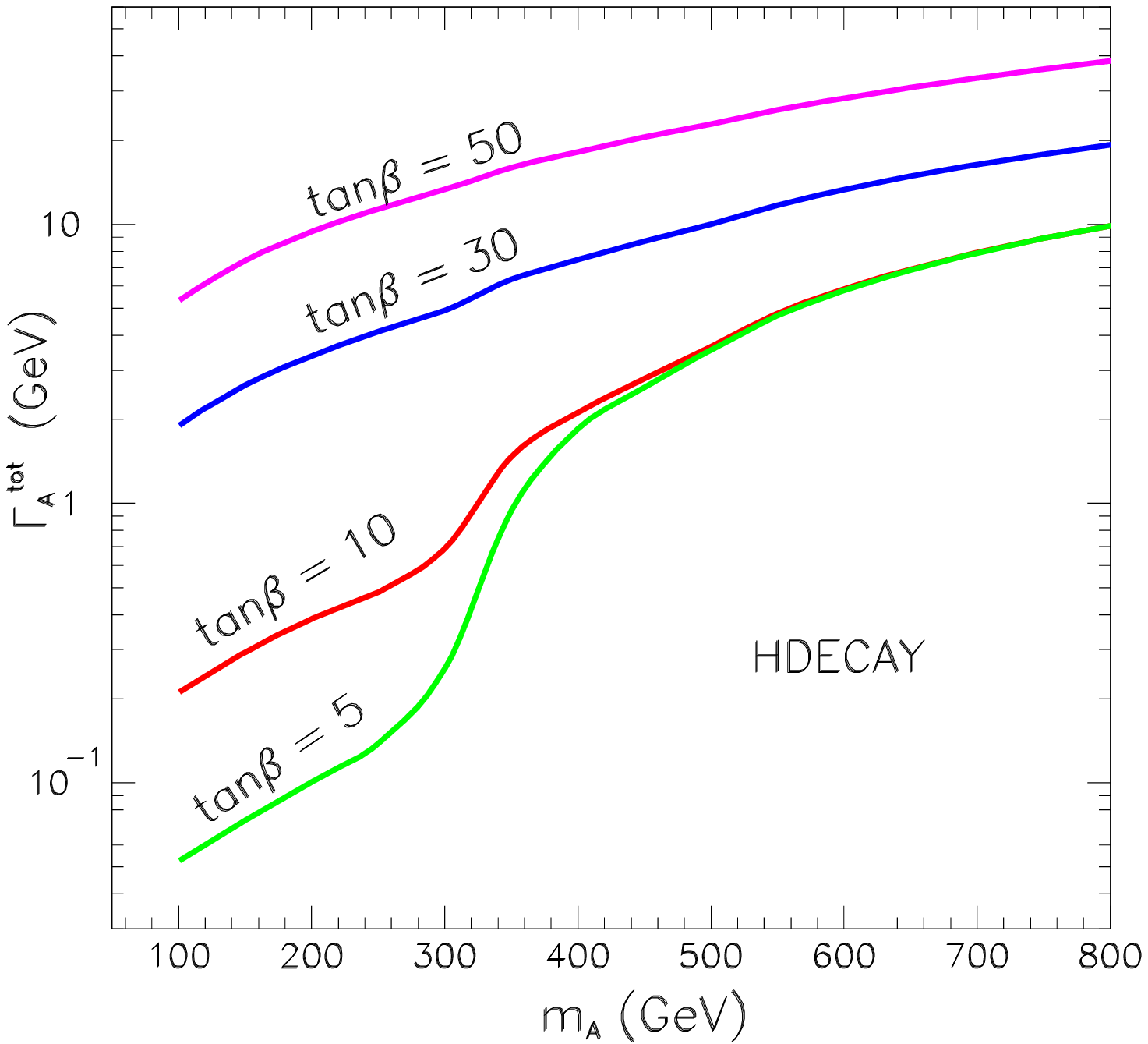}}
\end{center}
\caption {Total width of the $A^0$ boson as a function of $m_{A}$
for $tan\beta$ = 5 to $tan\beta$ = 50 calculated with HDECAY \cite{hdecay}. } 
\label{fig:awid}
\end{figure}

\begin{figure}[b]
\begin{center}
\resizebox{130mm}{100mm}{\includegraphics{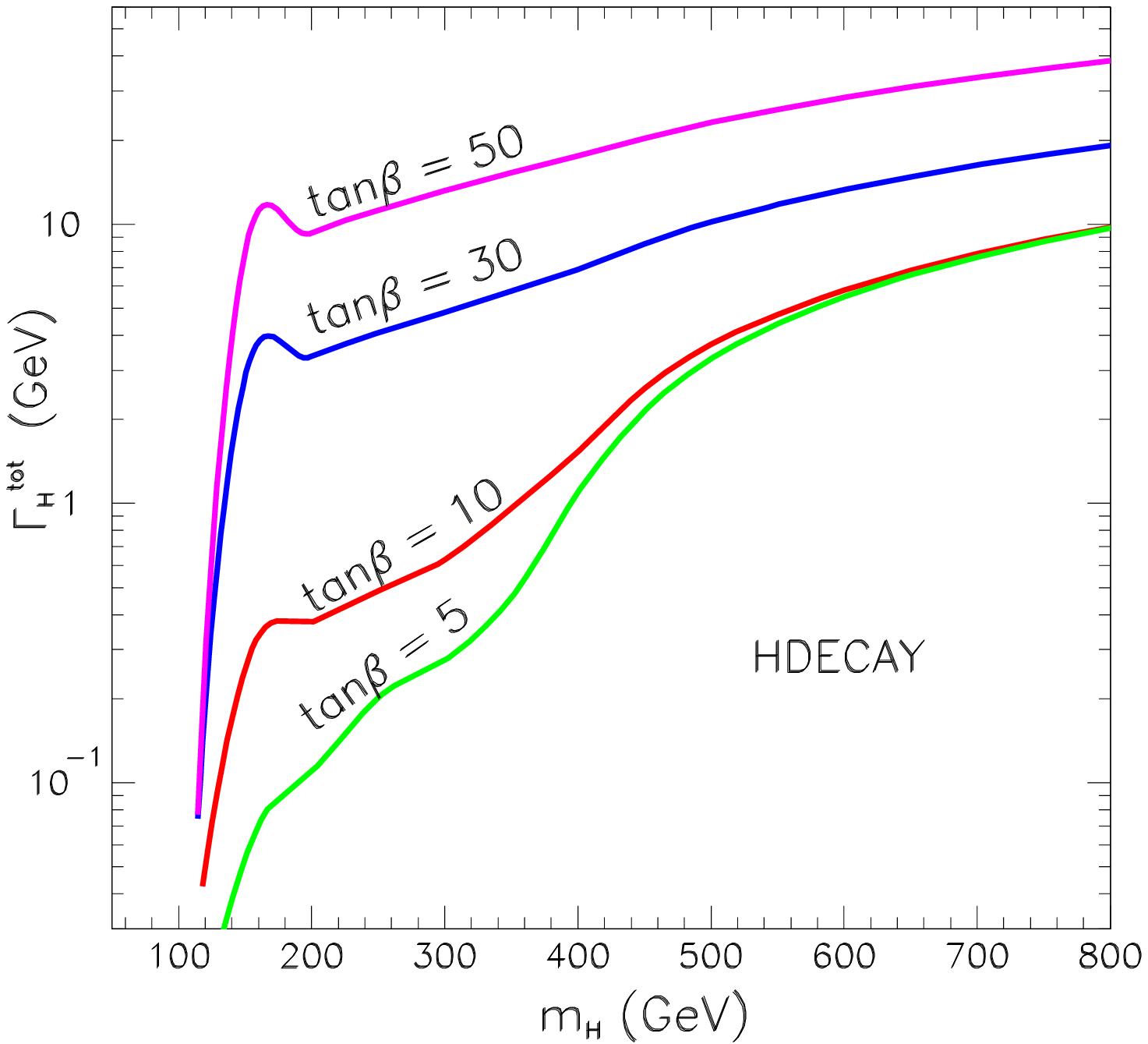}}
\end{center}
\caption {Total width of the $H^0$ boson as a function of $m_{H}$
for $tan\beta$ = 5 to $tan\beta$ = 50 calculated with HDECAY \cite{hdecay}.}
\label{fig:hwid}
\end{figure}

Figure \ref{fig:awid} shows the total width of the CP-odd MSSM Higgs boson $A$ as a function of $m_A$ for $tan\beta$ values between 5 and 50 calculated with HDECAY \cite{hdecay}. 
Figure \ref{fig:hwid} shows the same 
for the CP-even MSSM Higgs boson $H$. The total width
increases with increasing $tan\beta$ which in some cases may give the possibility to 
determine the value of $tan\beta$ from the width measurement. In CMS a sensitivity to 
two nearby states ($A$ and $H$) few~GeV apart, and possibly to the
Higgs width measurement could be expected in the $A, H \ra \mu\mu$ channel where 
effective mass resolution is at a percent level.

\subsection{Branching ratios and cross sections}

The branching ratios for $tan\beta$ = 5 and $tan\beta$ = 40 are shown in Figs. \ref{fig:br_hh_tanb5} and \ref{fig:br_hh_tanb40} for the heavy scalar $H$, in Figs. \ref{fig:br_a0_tanb5} and  \ref{fig:br_a0_tanb40} for the pseudoscalar $A$ and in Figs. \ref{fig:br_hl_tanb5} and  \ref{fig:br_hl_tanb40} for the light scalar $h$  calculated with the HDECAY \cite{hdecay} programs. 
Figures \ref{fig:br_hplus_tanb5} and \ref{fig:br_hplus_tanb40} show the same for
the charged Higgs. The branching ratios for the heavy
SUSY Higgs bosons are not sensitive to the amount of stop mixing and therefore the discovery
ranges calculated with the branching ratios shown in these figures are valid also for the 
maximal mixing scenario. These branching ratios are insensitive also to the sign of the
Higgsino mass parameter $\mu$, which according to the recent indirect measurements \cite{mupar},
is more likely to be positive. In general the PYTHIA 
branching ratios agree with those shown in the figures. In some cases, like for 
$h \ra \gamma\gamma$, there is a large disagreement between these two predictions 
and the HDECAY calculation is used.

The two most important production mechanisms for SUSY Higgs at the LHC are
the inclusive
gluon-gluon fusion $gg\rightarrow H_{SUSY}$ and the Higgs production
 in association with
b-quarks $gg\rightarrow b\bar{b}H_{SUSY}$. As the Higgs coupling to b-quarks (and to
$\tau$'s) is enhanced at high $tan\beta$ 
($g_{Hb\overline{b}}, g_{H\tau\tau} \sim cos\beta ^{-1}$)
 the associated production dominates at high
$tan\beta$ values and is about 90\% of the total rate for $tan\beta>$ 10 and
 $m_H \gsim$ 300~GeV. Thus the identification of b's 
and $\tau$'s is important for
Higgs searches at LHC implying the need for good tracking and impact parameter 
measurements \cite{tdr:tracker}. The gluon fusion
is mediated by quark loops, which can be affected by mixing.
Due to the dominance of the associated production and because only the CP-even Higgs 
can be affected, expectations for the heavy SUSY Higgs
are not sensitive to the loop effects unlike the inclusive $h \ra \gamma\gamma$
discussed in the following. 

\begin{figure}[t]
\begin{center}
\resizebox{140mm}{100mm}{\includegraphics{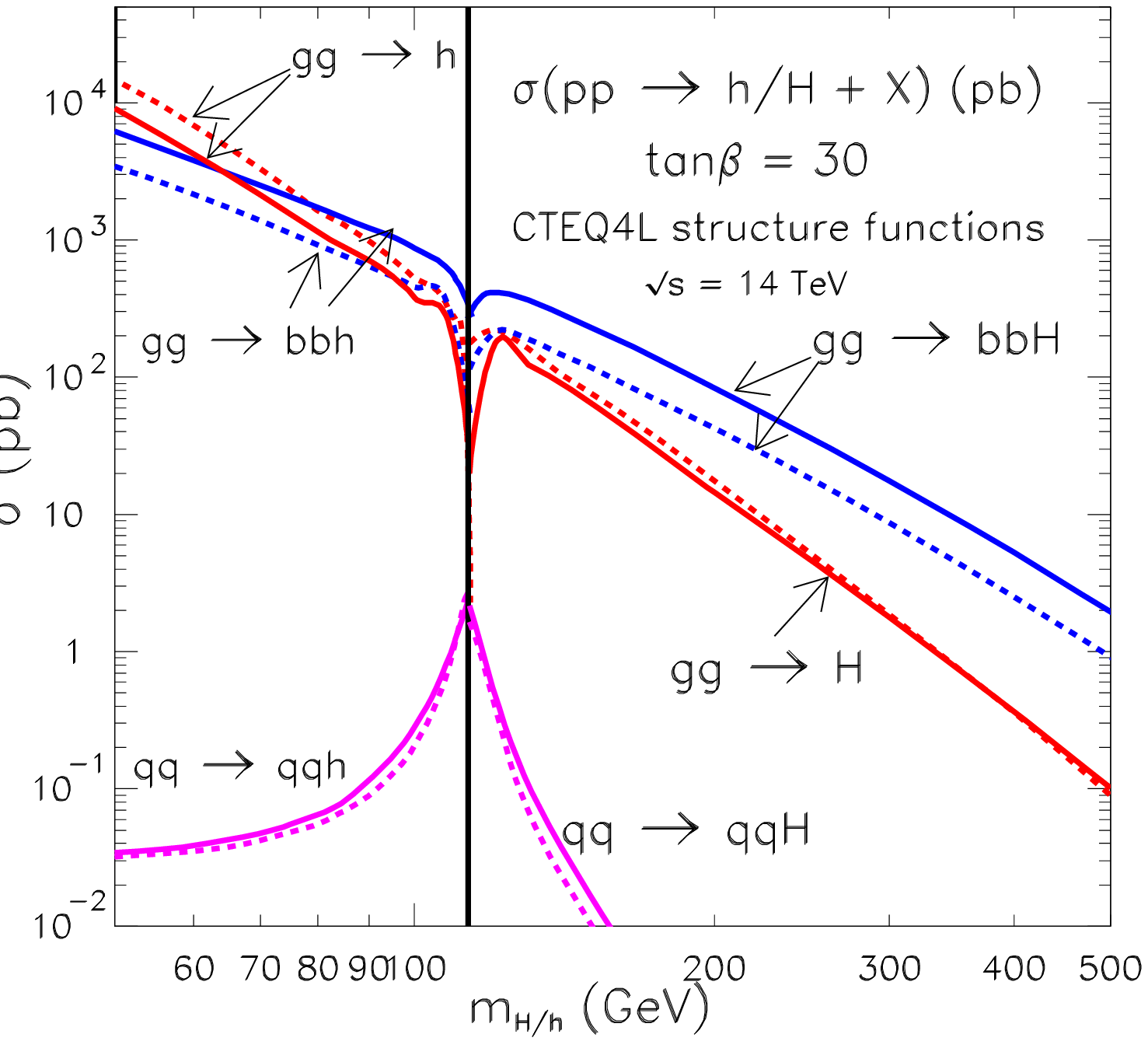}}
\end{center}
\caption {Cross sections for $gg \ra H,h$, $gg \ra b\overline{b}H,h$
and $qq \ra qqH,h$ as a function of the Higgs mass for $tan\beta$ = 30
with the CTEQ4L structure functions calculated with HIGLU/HQQ package \cite{higlu}.
The dashed lines are from PYTHIA6.1 with the same structure functions.}
\label{fig:sigmah}
\end{figure}

\begin{figure}[b]
\begin{center}
\resizebox{140mm}{100mm}{\includegraphics{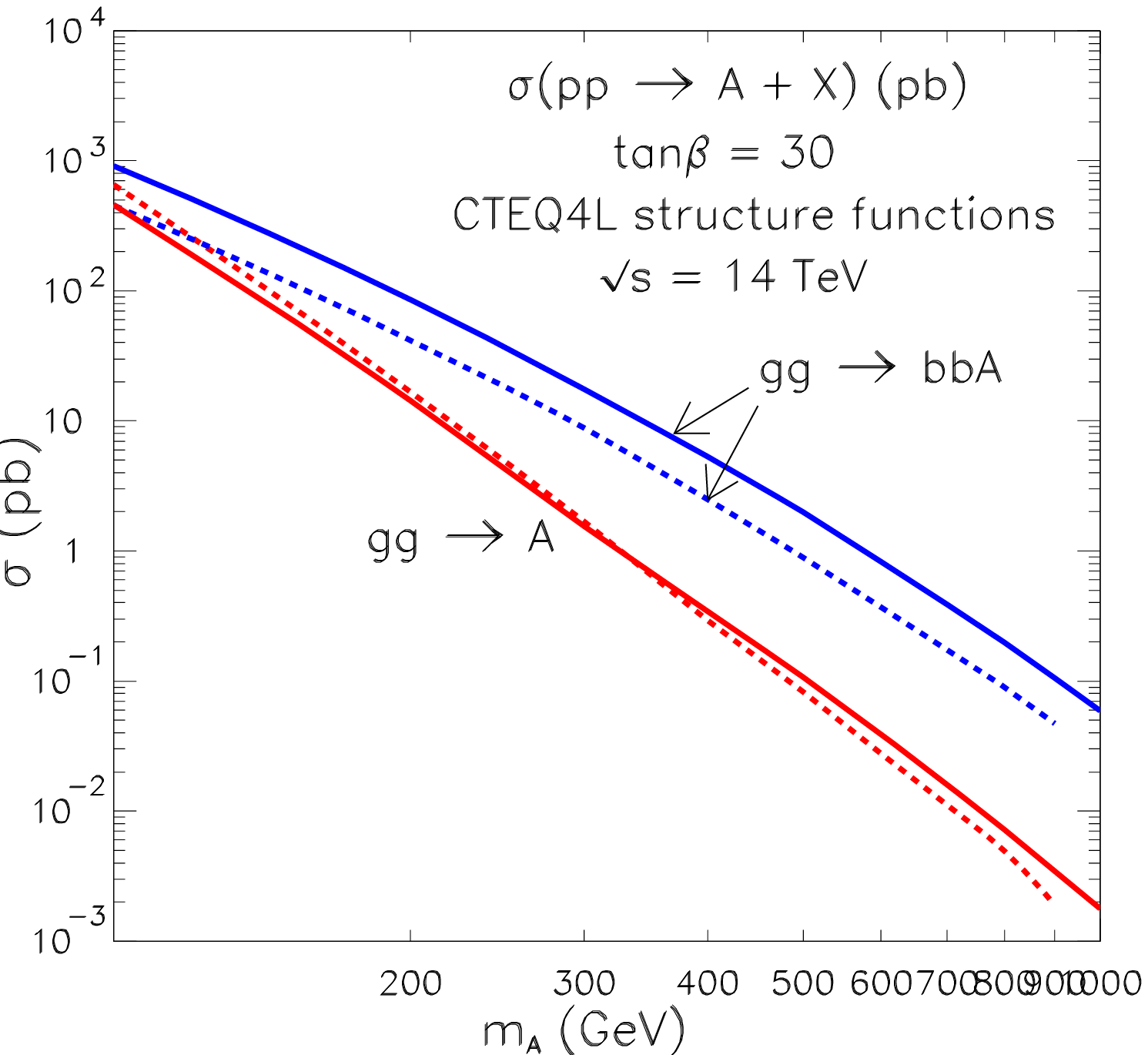}}
\end{center}
\caption {Cross sections for $gg \ra A$ and $gg \ra b\overline{b}A$
 as a function of the Higgs mass for $tan\beta$ = 30
with the CTEQ4L structure functions calculated with HIGLU/HQQ package \cite{higlu}.
The dashed lines are from PYTHIA6.1 with the same structure functions.}
\label{fig:sigmaa}
\end{figure}

\begin{figure}[t]
\begin{center}
\resizebox{140mm}{100mm}{\includegraphics{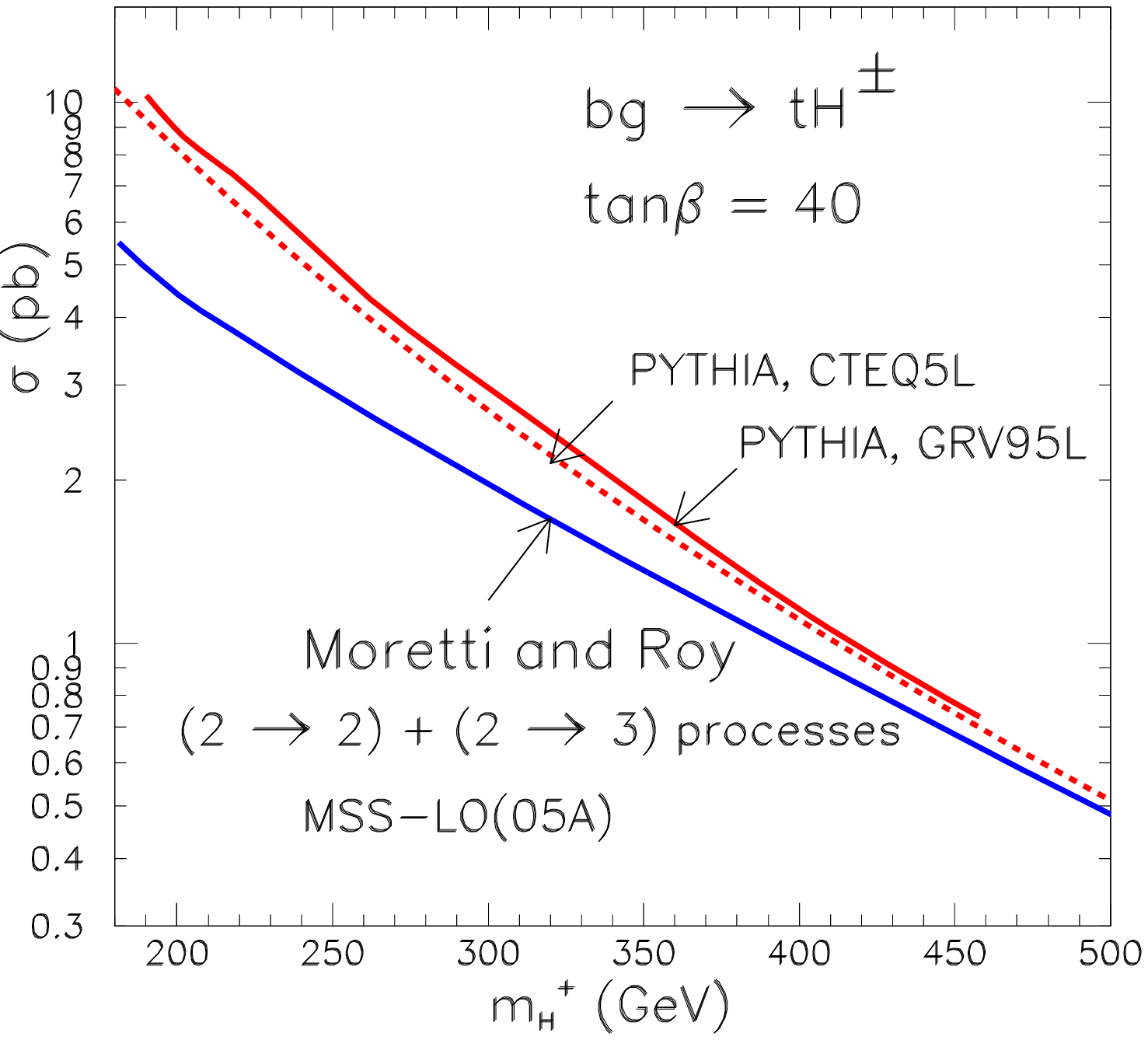}}
\end{center}
\caption {Cross section for $gb \ra tH^{\pm}$ for $tan\beta$ = 40
compared with the theoretical prediction of ref. \cite{moretti}.}
\label{fig:sigma_hplus}
\end{figure}

Figure \ref{fig:sigmah} shows the cross sections for the CP-even SUSY Higgs bosons $h$ and $H$  
as a function of the Higgs mass for $tan\beta$ = 30 with the CTEQ4L structure
functions calculated with the HIGLU \cite{higlu} and HQQ \cite{pphtt} programs. 
The cross sections are shown separately for the $gg \ra h/H$, $gg, qq \ra b\overline{b}h/H$
and $qq \ra qqh/H$ subprocesses. Figure \ref{fig:sigmaa} shows the corresponding predictions 
 for the CP-odd Higgs $A$. The 
cross sections are compared with results from 
PYTHIA calculated with the same structure functios. 
The PYTHIA points for $gg, qq \ra b\overline{b}H/A$ are systematically below the
theoretical
prediction indicating that in some cases our expectations may be conservative. 
The cross section for the charged Higgs in the process $bg \ra tH^{\pm}$ is shown
in Fig.\ref{fig:sigma_hplus} as a function of $m_{H^{\pm}}$ for $tan\beta$ = 40.
The result of the calculation of ref. \cite{moretti}, shown also in the figure,
includes both the 2 $\ra$ 2 and 2 $\ra$ 3 processes corrected for double counting.
 The PYTHIA prediction calculated using the process 
$bg \ra tH^{\pm}$ alone overestimates the cross section at low $m_{H^{\pm}}$ by 
a factor of $\sim$ 2.

\
\vspace{ 3mm}
\section{Processes}
\subsection{\boldmath{$h \ra \gamma\gamma$}}

The expected experimental sensitivity for the SM Higgs in the inclusive $H \ra \gamma\gamma$ channel is evaluated with full simulation of the CMS electromagnetic calorimeter and $\gamma\gamma$ mass reconstruction, including 
conversion recovery \cite{tdr:ecal,seez,lassila}. Figure \ref{fig:hgamma} shows the 
reconstructed Higgs mass superimposed on the total background for $m_H$ = 120~GeV and for 
100~$fb^{-1}$. The cross section times branching ratio required for a 5$\sigma$ significance for 100~$fb^{-1}$ is more than 40~$fb$ for zero stop mixing  ($m_h \lsim$ 113~GeV)  and more than 30~$fb$ for maximal stop mixing ($m_h \lsim$ 130~GeV). Higher cross sections are needed at lower mass values due to the increasing backgrounds. NLO cross sections are used for the signal and for the backgrounds.

\begin{figure}[t]
\begin{center}
\resizebox{130mm}{100mm}{\includegraphics{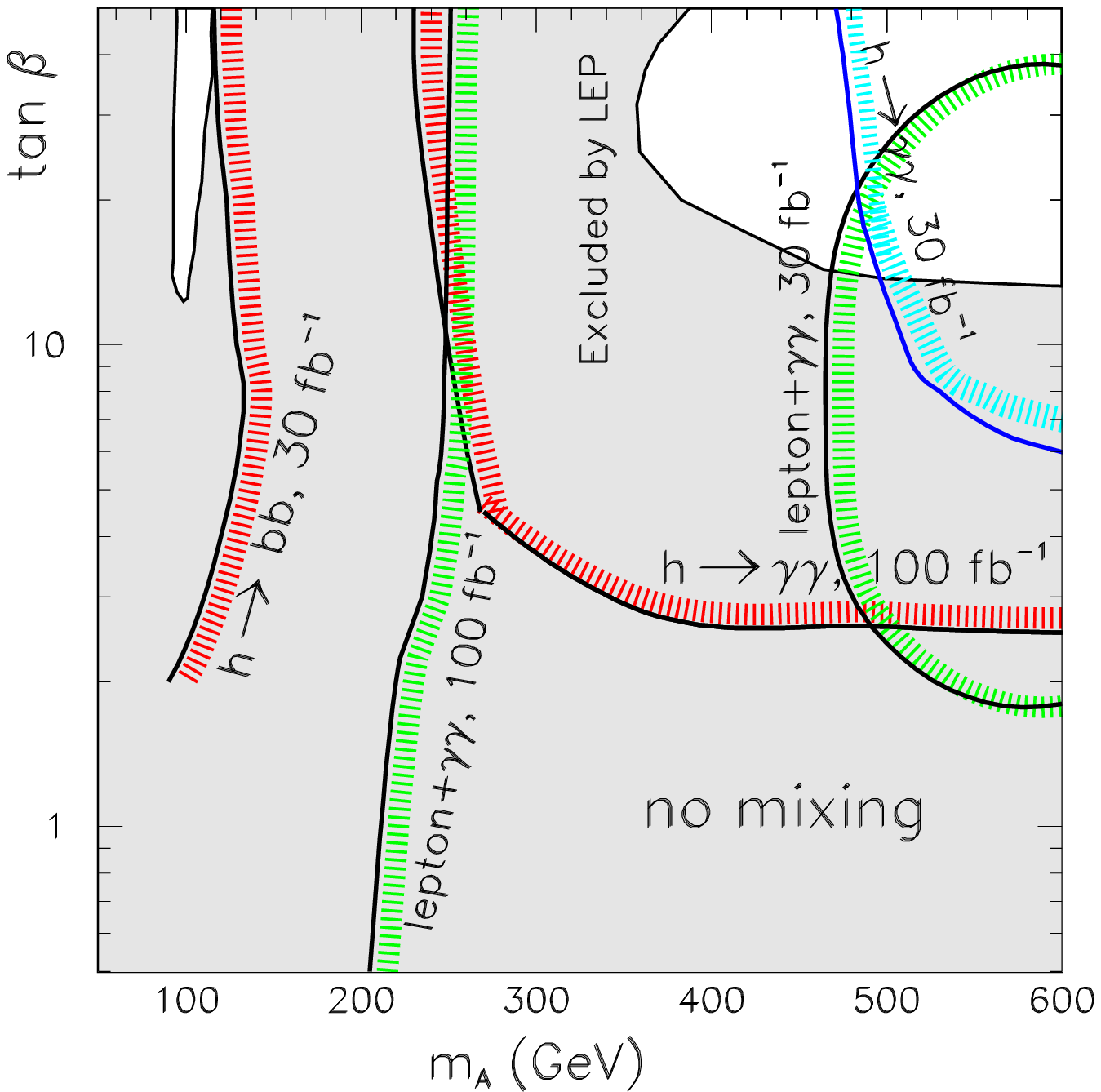}}
\end{center}
\caption {Expected 5$\sigma$ discovery reach for the inclusive 
$gg \ra h \ra \gamma\gamma$ for 30 and 100~$fb^{-1}$ 
and for $\ell\gamma\gamma$ in $t\overline{t}h$
and $Wh$ production and for $h \ra b\overline{b}$ in $t\overline{t}h$ production
for 30~$fb^{-1}$ as a function of $m_A$ and $tan\beta$ in the no-mixing scenario.
 The shaded area is excluded by LEP \cite{LEPII,junk}.} 
\label{fig:gamma_nomix}
\end{figure}

\begin{figure}[b]
\begin{center}
\resizebox{130mm}{100mm}{\includegraphics{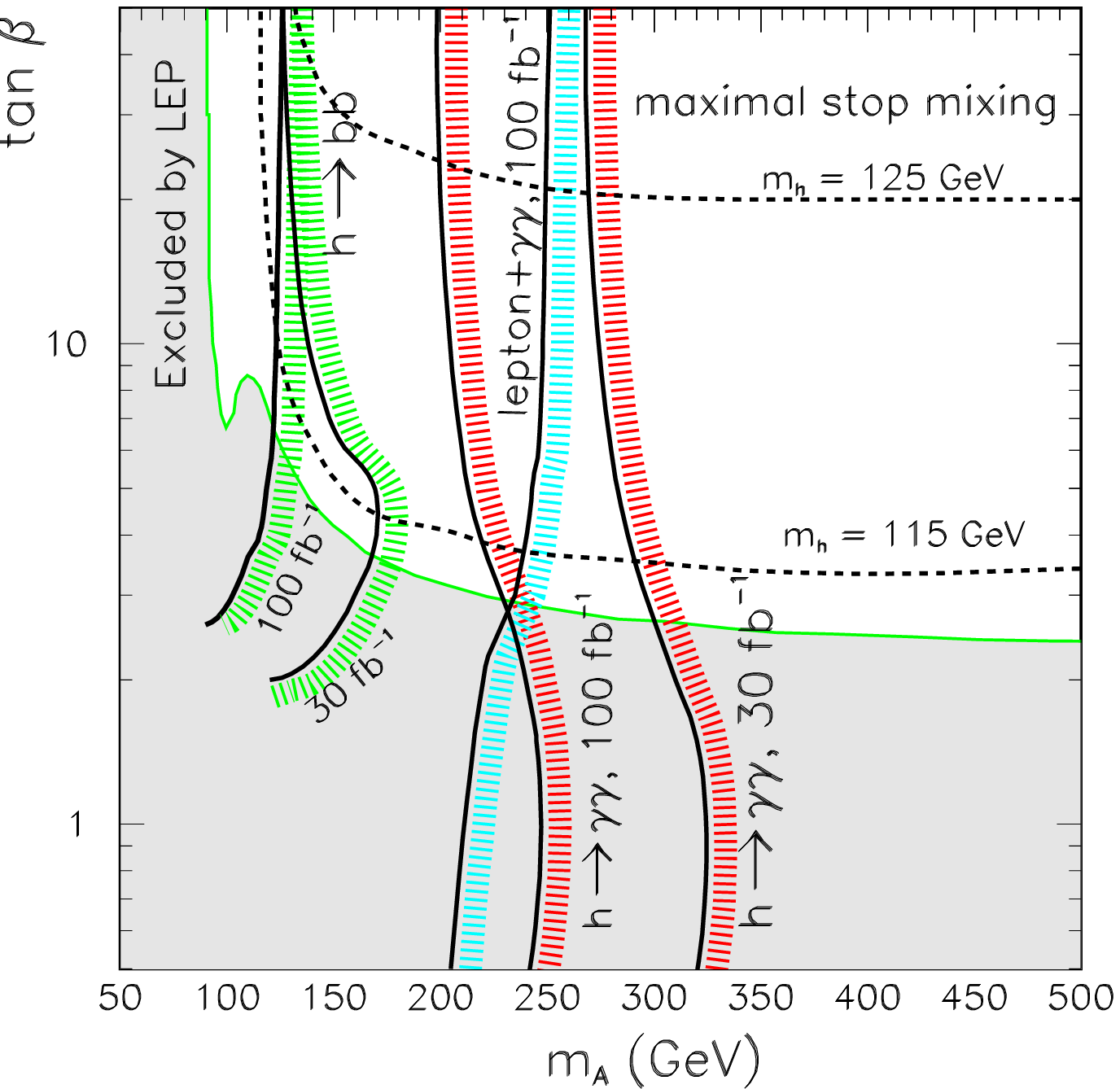}}
\end{center}
\caption {The same as in Fig. \ref{fig:gamma_nomix} but with maximal stop mixing. 
The dashed lines are the isomass curves for $m_h$ = 115~GeV and for $m_h$ = 125~GeV.} 
\label{fig:gamma_maxmix}
\end{figure}

The branching ratios for $h \ra \gamma\gamma$ are calculated with the HDECAY 
program \cite{hdecay} and the $gg \ra h$ cross section with the HIGLU package \cite{higlu} 
including the next-to-leading order 
corrections (NLO). Figure \ref{fig:gamma_nomix} shows the expected discovery range as a
function of $m_A$ and $tan\beta$ for 100~$fb^{-1}$ and for 30~$fb^{-1}$ 
assuming no stop mixing. The cross section is calculated using only the gluon-gluon fusion process, which gives a conservative limit. Figure \ref{fig:gamma_maxmix}
 shows the same in the maximal mixing scenario.
The parameter space excluded by LEP \cite{LEPII} data is also shown in the figures.

It has been shown in refs. \cite{djouadi,hep-ph/9806315} that the rate for 
$gg\ra h \ra \gamma\gamma$ could be strongly reduced in the case of large stop 
mixing if the stop becomes light, $m_{\tilde{t}_1}\lsim$200~GeV. 
The $\tilde{t}_1$ loop then
interferes destructively with the top quark loop partially cancelling each other, which
leads to a reduction of the $gg\ra h$ cross section. The $h\gamma\gamma$
coupling is also affected, but since the dominant contribution comes 
now from a $W$ loop, which interferes 
destructively with the top loop, a reduction of the top contribution by 
interfering stop loops increases the $h\ra\gamma\gamma$ partial width. 
However, this positive loop contribution is smaller 
than the negative for $gg\ra h$ and the net effect is a reduction of the overall
$gg\ra h \ra \gamma\gamma$ rate. 

\begin{figure}[t]
\begin{center}
\resizebox{130mm}{100mm}{\includegraphics{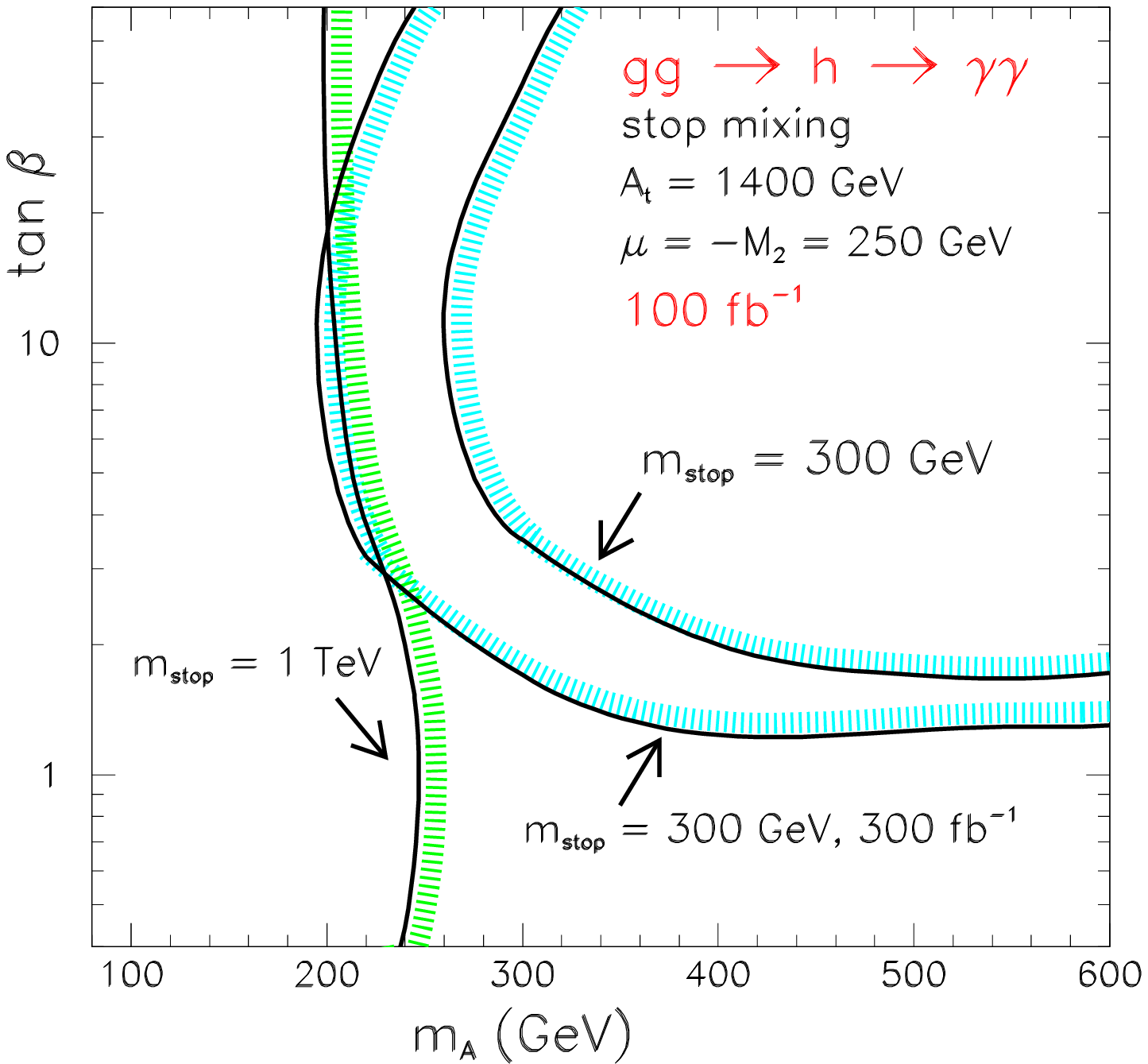}}
\end{center}
\caption {Expected 5$\sigma$ discovery reach for $gg \ra h \ra \gamma\gamma$
as a function of $m_A$ and $tan\beta$ with light stop, $m_{stop}$ = 300~GeV,
and large mixing, $A_t$ = 1400~GeV for 100 and 300~$fb^{-1}$ compared with the
expectation with a heavy stop for 300~$fb^{-1}$.} 
\label{fig:gamma_stop}
\end{figure}

The effect of light stop is studied using a modified HDECAY/HIGLU package \cite{hdecay}.
 Figure \ref{fig:gamma_stop}
shows the expected discovery range for $gg\ra h\ra \gamma\gamma$ assuming large mixing, $\tilde{A}_t$ = 1400~GeV, and a light stop, $m_{\tilde{t}_1}$ = 300~GeV as a
function of $m_A$ and $tan\beta$ for 30~$fb^{-1}$ and for 100~$fb^{-1}$
 \cite{loops} compared with our standard expectation with heavy stop.
The SUSY parameters $\mu$ and $m_2$ are taken to be $\mu$ = -250 and $M_2$ = 250~GeV.
 For $m_{\tilde{t}_1}$ = 200~GeV the reduction is so large that no discovery would be possible through the gluon fusion channel $gg\ra\ h \ra\gamma\gamma$. 
The only alternatives are then the tree-diagram associated production channels 
$Wh$ and $t\overline{t}h$.

An excellent ECAL resolution is mandatory for the inclusive  $h \ra\gamma\gamma$
channel due to the large prompt $\gamma\gamma$ background ($S/B \sim$ 1/10). 
The search of 
$h \ra\gamma\gamma$ in the associated production channels $Wh$ and $t\overline{t}h$ 
 is less sensitive to the $\gamma\gamma$ mass
resolution as in this case the backgrounds can be effectively reduced by the 
lepton requirement giving $S/B \sim$ 1/1. However the event rate is small
 making these channels useful only at
the ultimate luminosities, $Lt \gsim 100~fb^{-1}$. These channels are studied in ref. 
\cite{mile} with the fast detector simulation method. The expected discovery reach for 
$Lt \gsim 100~fb^{-1}$ may exceed that for the inclusive channel at low $tan\beta$ 
values, as is shown in Figs. \ref{fig:gamma_nomix} and \ref{fig:gamma_maxmix}. The 
importance of the $Wh$ and $t\overline{t}h$ channels is not only that they could be 
$H(h)$ discovery channels in case of suppressed $gg \ra h$ production or difficulties
in monitoring at a $\lsim$ 1\% level the electromagnetic calorimeter 
calibration, but
also because they provide measures of the $WWH$ and $t\overline{t}H$ couplings.

\subsection{\boldmath{$h \ra b\overline{b}$}}

The $h \ra b\overline{b}$ decay channel benefits from a large branching ratio
over the whole $m_A$-$tan\beta$ parameter space, the same is true for the
SM Higgs for $m_H \lsim$ 130~GeV. A detailed study of the $H \ra b\overline{b}$ 
decay channel in $WH$ and $t\overline{t}H$ events is presented in ref. 
\cite{volker}
and in $t\overline{t}h$ events in refs. \cite{volker2,dangreen}.
To extract the Higgs signal in $t \overline{t} H \ra l^\pm \nu q \overline{q} 
b \overline{b} b \overline{b}$ events requires tagging of up to
4 b-jets in the presence of a large hadronic activity, 
reconstruction of the Higgs mass from two b-jets and the reconstruction
of the associated leptonic and hadronic top or the W. A sophisticated likelihood
method has been developed in ref. \cite{volker} to optimise $H \ra b\overline{b}$ 
signal visibility in the $t\overline{t}H$ channel. The main backgrouds 
($t\overline{t}b\overline{b}$, $t\overline{t}jj$, $t\overline{t}Z$) are generated
 with the CompHep-PYTHIA package \cite{CompHep,CPinterf} which includes 
the calculation of the matrix elements with higher
order corrections. Figure \ref{fig:hbb} shows the 
reconstructed invariant mass for the SM Higgs boson 
superimposed on the total background for $m_H$ = 115~GeV 
and for 30~$fb^{-1}$. The $t\overline{t}H$ channel
is far more promising than the $WH$ channel. The signal 
to background ratio in $t\overline{t}H$ is of the
order of one and the discovery range is $m_H \lsim$ 122~GeV 
for 30~$fb^{-1}$ (signal visibility degrades with increasing mass).
 Figures \ref{fig:gamma_nomix} and \ref{fig:gamma_maxmix}
 show the expected 
discovery ranges for the MSSM Higgs in $t\overline{t}h$ final states in the 
no-mixing and in the maximal mixing (maximal $m_h$) scenario
for 30~$fb^{-1}$ and for 100~$fb^{-1}$. For observation of $H \ra 
b\overline{b}$ in the $WH$ channel, ultimate luminosities, in excess of $\gsim 200~fb^{-1}$, are needed due to the much lower signal to
background ratio ($\sim$ 1/40). The discovery reach for 300~$fb^{-1}$ 
is at $m_H \lsim$ 123~GeV 
using lowest order calculations (k=1). The $m_A$-$tan\beta$ discovery
contours for the $Wh$ channel are similar to those for the $t\overline{t}h$,
 but about ten times more integrated luminosity is necessary to 
explore the same amount of parameter space.
The importance of the $WH(Wh)$ channel is that it could provide a measure
of the $WWH$ coupling.

\subsection{\boldmath{$A, H \ra \mu\mu$}}

Although the branching ratio for $A, H \ra \mu\mu$ is small, $\sim 3 \times 10^{-4}$,
the associated $b\overline{b}H_{SUSY}$ production is enhanced at large $tan\beta$. This channel is very interesting as it allows a precise reconstruction of the Higgs 
boson mass thanks to the excellent muon momentum resolution of CMS \cite{tdr:muon}. 
$A, H \ra \mu\mu$ is studied in ref. \cite{furic} using for the muon momentum resolution 
results from a full simulation of the CMS tracker. The expected Higgs mass resolution
 for $m_A$ = 150~GeV ($m_H$ = 151.2~GeV) at $tan\beta$ = 15, for instance, is 2.2~GeV 
with superimposed unresolved CP-even ($H$) and CP-odd ($A$) states. 
B-tagging greatly improves signal
visibility at large $tan\beta$ reducing the overwhelming $Z,\gamma^* \ra \mu\mu$ background. The expected discovery
 reach for 100~$fb^{-1}$ is shown in Fig.\ref{fig:maxmix_100fb}. At large $tan\beta$
the natural width of $A$ and $H$ for accessible masses is in the few~GeV to $\sim$ 10~GeV range (Figs. \ref{fig:awid}, \ref{fig:hwid}). With a $\sim$ 1\% level $\mu\mu$ mass
resolution a fit to the $\mu\mu$ signal shape would allow measurement of its natural width.

\subsection{\boldmath{$A, H \ra \tau\tau$ with $2~lepton$, $lepton+\tau~jet$ 
and $2~\tau~jet$ final states}}

A systematic study of the $A, H \ra \tau\tau$ decay with $2~lepton$, 
$lepton+\tau~jet$ and $2~\tau~jet$ final states is presently in progress in CMS including full simulation
of the hadronic $\tau$ trigger, $\tau$ identification, $\tau$ tagging with 
impact parameter and vertex reconstruction, Higgs mass reconstruction and 
b-tagging in the associated production channels.

 A detailed study of the $A, H \ra \tau\tau$ with $e+\mu$ and $\ell^+\ell^-$
final states is presented in ref. \cite{sami}. Tagging of the associated b-jets
and the impact parameter tagging of the two $\tau$'s are investigated with CMSIM 
simulation. The associated b-jets are soft and uniformly distributed in the 
central and endcap areas of the CMS tracker. The study shows that a b-tagging
 efficiency of $\sim$ 35\% can be obtained for these b-jets keeping the 
mistagging rate in the $Z+jets$ events under 1\% level. Promising 
results are also obtained for $\tau$ tagging through
 impact parameter measurements
of the two hard leptons from the $\tau$ decays. Tau tagging is needed to 
suppress the backgrounds where
the leptons originate from W or Z. This method allows to double the signal statistics
 by including all $\ell^+\ell^-$ and not only $e+\mu$
final states. The expected discovery range is shown in Figs. \ref{fig:nomix_30fb} 
and \ref{fig:maxmix_30fb} for 30~$fb^{-1}$ and in Fig. \ref{fig:maxmix_100fb}
 for 100~$fb^{-1}$ separately
for the $e+\mu$ and $\ell^+\ell^-$ final states. Examples of the $e + \mu$ - and
$\ell^+\ell^-$ signals are shown in Figs. \ref{fig:hmass_emu} and \ref{fig:hmass_ll}.  

$A, H \ra \tau\tau$ with 2~$\tau~jet$ hadronic final sates have been shown to extend
significantly the SUSY Higgs discovery reach into the large mass (600 - 800~GeV)
range \cite{h2jet}. To exploit fully the
$2~\tau~jet$ final states - especially in the low ($\sim$ 200~GeV) mass range -
 an efficient hadronic $\tau$ trigger has been developped 
based on Level-1 calorimeter selection, Level-2 electromagnetic
calorimeter isolation and collimation \cite{trigger1} and a Level-3 tracking 
(isolation) in the pixel detector \cite{trigger3}.
A reduction factor of $\sim$ 10$^3$ against the QCD background is obtained
with the full High Level Trigger selection (Level2 and Level3) with
an efficiency of $\sim$ 35\% for the signal at $m_H$ = 200 and 500~GeV \cite{sasha4}.

Offline $\tau$ identification in the hadronic final states is based on a requirement of
an isolated hard ($p_t >$40~GeV) charged track inside the jet. This $\tau$ selection
gives a rejection factor of $\sim 3 \times 10^{-4}$ against a QCD jet with $E_t \sim$
100 - 200~GeV. A significant
improvement can be expected from the impact parameter measurement 
of the two hard tracks as discussed above in the case of the $\ell^+\ell^-$ final states. 
Furthermore, including the 
$\tau$ to 3-prong decays in a small cone in the core of the calorimetric 
jet, the event rate for
$A, H \ra \tau\tau \ra 2~\tau~jets$ is enhanced by a factor of $\sim$1.7. 
The rejection factor against the QCD jets degrades by a factor of $\sim$ 3 when 
the 3-prong decays are included. 
However this background may also be suppressed by secondary vertex reconstruction.
Promising results are obtained from 
a study on the identification of the 
3-prong $\tau$ vertex with full (ORCA) track reconstruction \cite{tau_vertex}. A
rejection factor of $\sim$ 5 is obtained against the 3-prong QCD jets 
with an efficiency of $\sim$ 70\% for the $\tau$ jets.     

The resolution of the reconstructed Higgs mass and even more so the mass reconstruction
efficiency in $A, H \ra \tau\tau$ events is very sensitive 
to the $E_t^{miss}$ measurement. The absolute value of $E_t^{miss}$  is relatively
small in these events making the mass reconstruction 
and background reduction with 
a cut in $E_t^{miss}$ a difficult task. Most recent results from full simulation 
\cite{sasha} indicate a good mass resolution and reconstruction efficiency 
and confirm the earlier results of the fast simulation study \cite{h2jet}.  
The expected discovery reaches for the $2~\tau~jet$ and the $lepton+\tau~jet$ \cite{hljet} final states 
are shown in Figs. \ref{fig:nomix_30fb} (no mixing) and \ref{fig:maxmix_30fb} (maximal mixing) for 30~$fb^{-1}$.

\subsection{\boldmath{$H^{\pm} \ra \tau\nu$ in $gb \ra tH^{\pm}$ and in $q\overline{q'} \ra H^{\pm}$}}

Search for the charged Higgs at LHC is essential for the understanding
 of the nature of the Higgs sector. $H^{\pm} \ra \tau\nu$ with a hadronic
 $\tau$ decay has been shown to lead to the 
most favourable signature. The signal can be strongly enhanced against the
background from $W \ra \tau\nu$ decays exploiting the $\tau$ polarization in 
the one-prong $\tau \ra \pi^{\pm} (\pi^0) \nu$ decays \cite{dproy}. 
Due to the $\tau$ polarization, the single
pion from a $\tau$ decay is harder when the $\tau$ originates from an $H^{\pm}$ than
from a $W$. Requiring 80\% of the visible $\tau$-jet energy to
be carried by the single pion, the $t\overline{t}$ background can be reduced by
a factor of $\sim$ 300 while keeping the signal efficiencies at a 10\% to 20\% level 
(including the jet $E_t$ threshold) \cite{hplus}.
 
\begin{figure}[b]
\begin{center}
\resizebox{130mm}{100mm}{\includegraphics{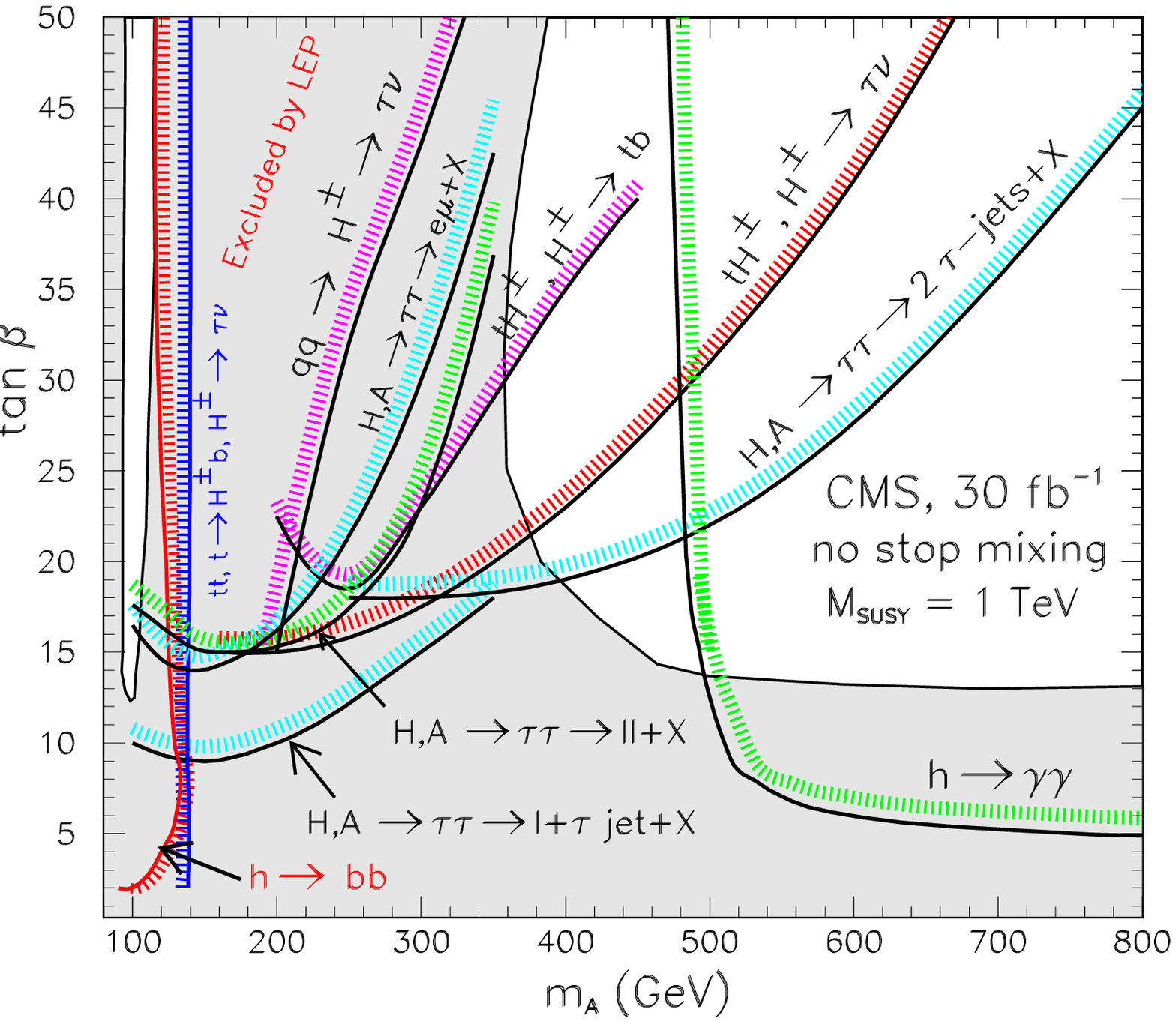}}
\end{center}
\caption {Expected 5$\sigma$ discovery reach for the MSSM Higgs bosons in 
CMS in the no-mixing scenario with 30$fb^{-1}$
as a function of $m_A$ and $tan\beta$. The shaded area is excluded by LEP \cite{LEPII,junk} } 
\label{fig:nomix_30fb}
\end{figure}

\begin{figure}[t]
\begin{center}
\resizebox{130mm}{100mm}{\includegraphics{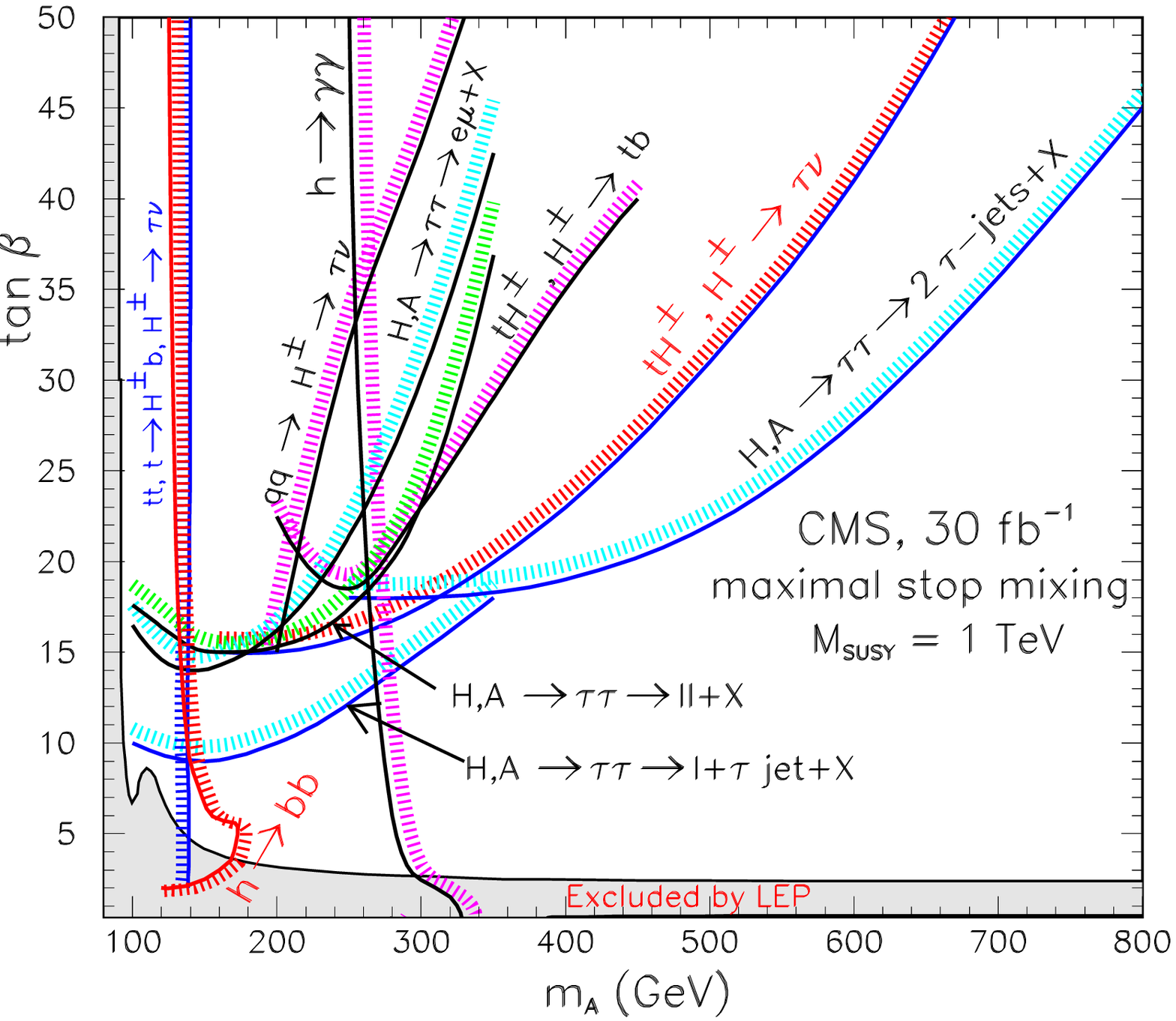}}
\end{center}
\caption {Expected 5$\sigma$ discovery reach for the MSSM Higgs bosons in 
CMS in the maximal mixing scenario for 30$fb^{-1}$
as a function of $m_A$ and $tan\beta$. The shaded area is excluded by LEP \cite{LEPII,junk}.} 
\label{fig:maxmix_30fb}
\end{figure}

 If the charged 
Higgs is light, $m_{H^+} < m_{top}$, the production is through the 
  the $t\overline{t}$ events followed by $t \ra H^{\pm}b$ and the
 discovery range is limited by the top mass 
\cite{manas}. The expected discovery range for 30~$fb^{-1}$ is 
for $m_A \lsim$ 160~GeV almost
independent of $tan\beta$ and is shown in Fig. \ref{fig:nomix_30fb}. 
 For a heavier charged Higgs, $m_{H^+} > m_{top}$,
 the main production processes are $gg \ra tbH^{\pm}$,
$gb \ra tH^{\pm}$ and $q\overline{q} \ra H^{\pm}$; the two first processes
partially overlap with a b from the proton structure. The production in
 association with the top quark is shown to lead 
to the most favourable final states if purely hadronic final states $t \ra qqb$
are selected, with an almost background-free signal in the 
transverse mass reconstructed from the $\tau$ jet and the transverse missing energy
\cite{dproy,hplus}. Figure \ref{fig:hplus_mass} shows the 
reconstructed transverse mass for $m_{H^{\pm}}$ = 217~GeV 
($m_A$ = 200~GeV) and $tan\beta$ = 40 
superimposed on the total background for 30$fb^{-1}$. 
Determination of the Higgs mass from the endpoint may be possible with a $\sim$ 2\% 
precision. Results of ref. \cite{hplus} are obtained using the PYTHIA6.1 cross 
sections and branching ratios which are found to be in a rough agreement with the 
theoretical expectations \cite{moretti}. 
The expected discovery reach is shown in  Figs. \ref{fig:nomix_30fb} 
for 30~$fb^{-1}$ and in Fig. \ref{fig:maxmix_100fb} for 100~$fb^{-1}$.

The s-channel production of $H^{\pm}$ in $q\overline{q}' \ra H^{\pm} \ra \tau\nu$ is investigated using again the hadronic $\tau$ decay and exploiting 
the $\tau$ polarization 
method discussed above \cite{serguei}. It is difficult to obtain
a signal as the reconstructed $H^{\pm}$ transverse mass 
 is on the tail of the very large $q\overline{q}' \ra W \ra \tau\nu$ background. 
It is shown in ref. \cite{serguei} that nevertheless the Higgs mass and 
$tan\beta$ may be still
extracted using fits to the transverse mass distributions. The expected discovery reach is shown in Figs. \ref{fig:nomix_30fb} 
and \ref{fig:maxmix_30fb} for 30~$fb^{-1}$.

\subsection{\boldmath{$H^{\pm} \ra tb$ in $gb \ra tH^{\pm}$}}

The other important decay channel of the charged Higgs, $H^{\pm} \ra tb$, 
has been investigated recently in $gb \ra tH^{\pm}$ production, requiring one
isolated lepton from the decay of one of the top quarks \cite{nikita}.  To extract
the Higgs signature in these multijet events requires tagging of three b-jets,
reconstruction of the leptonic and hadronic top quark and the Higgs mass reconstruction
from a top quark and one b-jet. The study is made with a realistic b-tagging simulation  based on the CMSIM reconstruction of the CMS tracker, and taking into
account the correct kinematics of the jets in the event. After selection cuts 
and b-tagging, the background is concentrated in the signal area making the identification
of the Higgs mass peak difficult in this channel.
The expected discovery range at high $tan\beta$ is shown in Fig.
\ref{fig:nomix_30fb}  for 30~$fb^{-1}$.

\subsection{\boldmath{$A, H \ra \chi^{0}_{2}\chi^{0}_{2}  \ra 4\ell^{\pm}+X$}}

\begin{figure}[b]
\begin{center}
\resizebox{130mm}{100mm}{\includegraphics{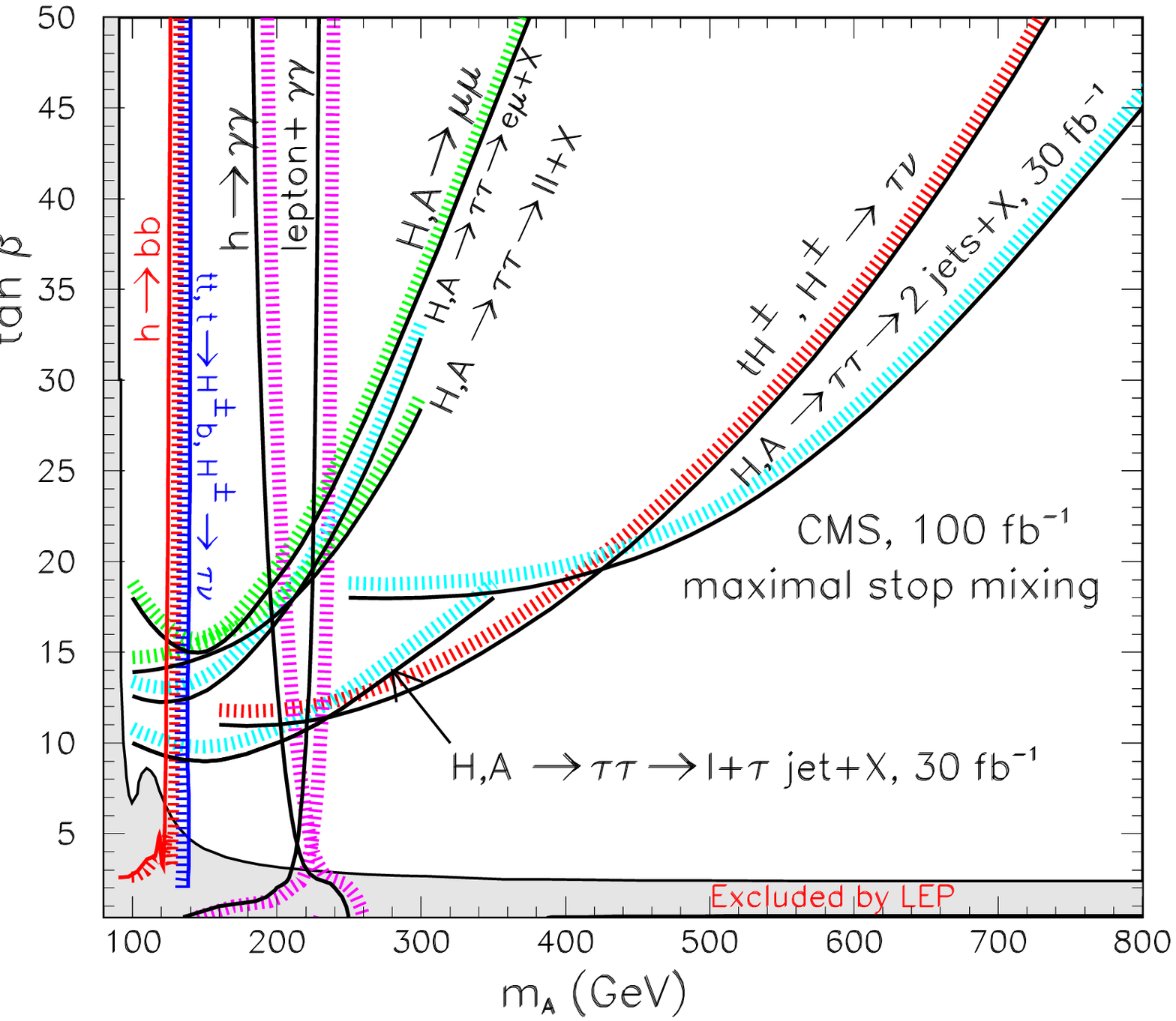}}
\end{center}
\caption {The same as in Fig. \ref{fig:maxmix_30fb} but for 100$fb^{-1}$. 
The discovery reaches for $H,A \ra \tau\tau \ra 2~\tau~jets$ 
and for $H,A \ra \tau\tau \ra lepton+\tau~jet$ are shown for 30$fb^{-1}$.
 The shaded area is excluded by LEP \cite{LEPII,junk}} 
\label{fig:maxmix_100fb}
\end{figure}

\begin{figure}[t]
\begin{center}
\resizebox{130mm}{90mm}{\includegraphics{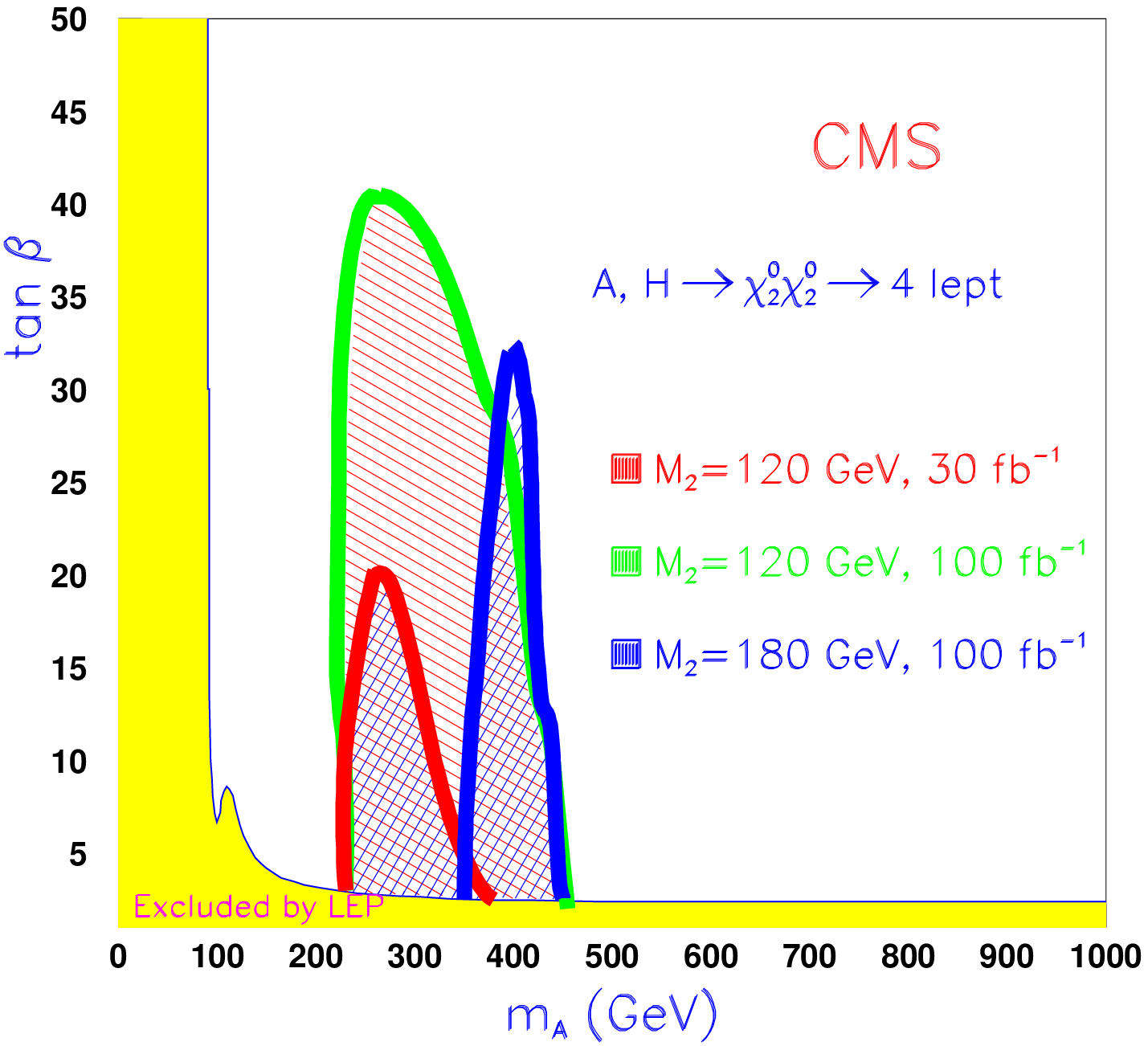}}
\end{center}
\caption {Expected 5$\sigma$ discovery reach for $H,A \ra \chi^0_2 \chi^0_2 \ra 4l^{\pm}$
as a function of $m_A$ and $tan\beta$ with $M_1$ = 60~GeV, $M_2$ = 120~GeV, $\mu$ = -500~GeV, 
$M_{\tilde{\ell}}$ = 250~GeV, $M_{\tilde{q},\tilde{g}}$ = 1000~GeV, for 30 and 100~$fb^{-1}$. The discovery reach assuming $M_2$ = 180~GeV is
also shown for 100~$fb^{-1}$.} 
\label{fig:chi2chi2}
\end{figure}

\begin{figure}[b]
\begin{center}
\resizebox{130mm}{100mm}{\includegraphics{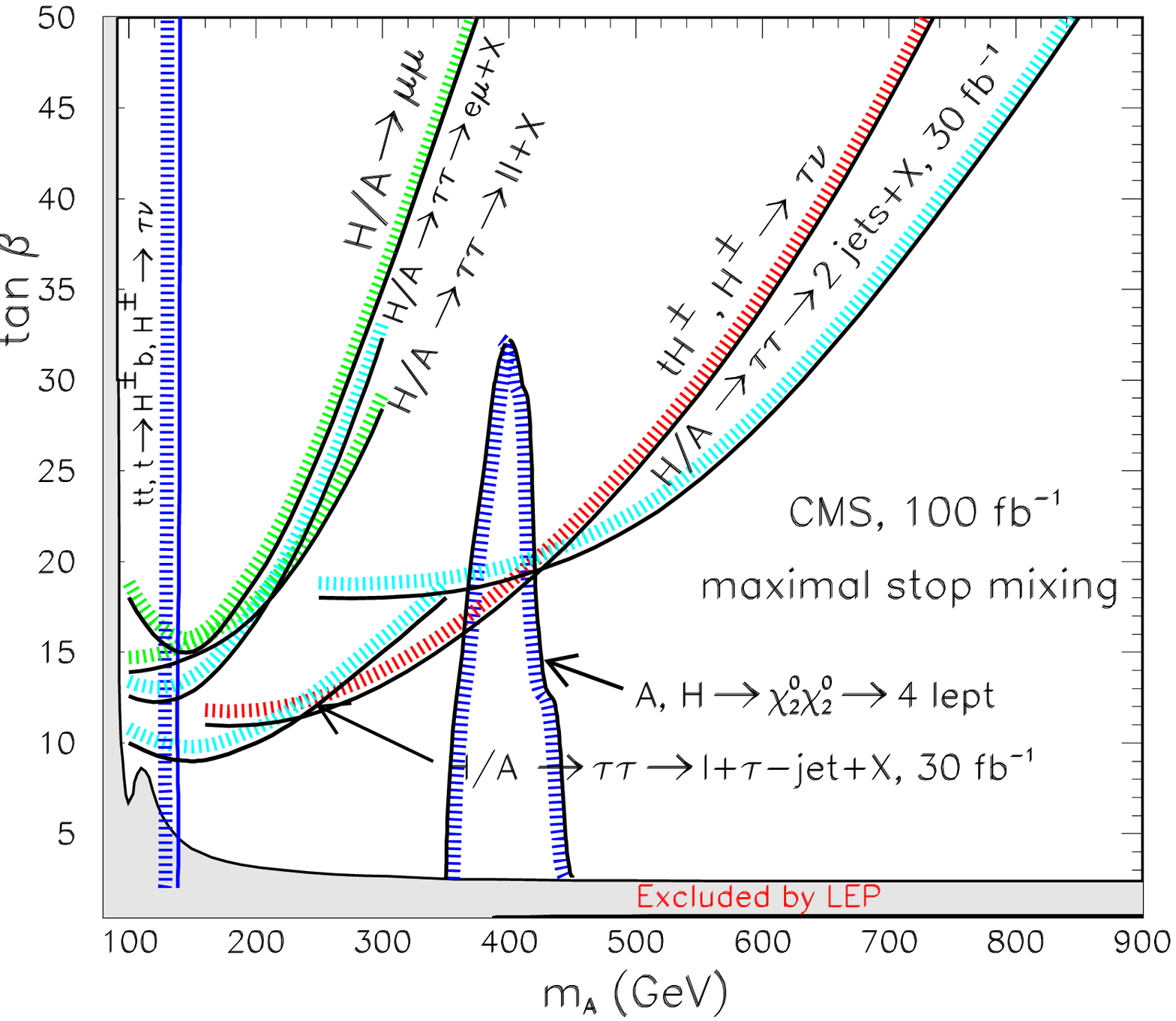}}
\end{center}
\caption {Expected 5$\sigma$ discovery reach for the heavy MSSM 
Higgs bosons including $H,A \ra \chi^0_2 \chi^0_2 \ra 4l^{\pm}$ 
as a function of $m_A$ and $tan\beta$ for 100$fb^{-1}$ assuming $M_1$ = 90~GeV, 
$M_2$ = 180~GeV, $\mu$ = 500~GeV,
$M_{\tilde{\ell}}$ = 250~GeV, $M_{\tilde{q},\tilde{g}}$ = 1000~GeV.} 
\label{fig:maxmix_100fb_susy}
\end{figure}

As can be seen from Figs. \ref{fig:nomix_30fb} - \ref{fig:maxmix_100fb}, 
 essentially the whole MSSM $m_A$-$tan\beta$ 
parameter space not excluded by LEP 
is expected to be covered by the light SUSY Higgs signal with $h \ra \gamma\gamma$
and $h \ra b\overline{b}$ decays in both the no-mixing and maximal $m_h$ scenarios
 already with 30~$fb^{-1}$.
At high $tan\beta$ the heavy SUSY Higges are
expected to be found with the decays to $\tau$'s and muons, the discovery ranges 
extending down to $tan\beta \sim$ 10 for the low mass range $m_A \lsim$ 200~GeV and 
to $tan\beta \sim$ 20 for the high mass range $m_A \gsim$ 300~GeV. 
If $tan\beta$ is 
between these boundaries and the LEPII limit it may be difficult to identify the 
nature (SM or SUSY) of the only $h \ra \gamma\gamma$ or $h \ra b\overline{b}$
discovered Higgs state. It is shown in ref. \cite{filip} that in this 
part of the $m_A$-$tan\beta$ parameter space Higgs decays
 to sparticles (Figs. 3-5) can be used to complete
the search with SM particle decays. The channel 
$A, H \ra \chi^{0}_{2}\chi^{0}_{2}  \ra 4\ell^{\pm}+X$ turns out to be the most 
favourable one provided neutralinos and sleptons are light enough so that the 
$\chi^{0}_{2} (\ra \tilde{\ell}\ell) \ra \chi^{0}_{1}\ell^+\ell^-$ branching ratio
is significant. The four isolated lepton final state signature with little additional jet activity allows to reduce drastically the background levels.   
For this channel the backgrounds coming from other 
SUSY processes dominate and have to be taken into account.
 Figure \ref{fig:filip_mass} shows the 
signal in the 4-$lepton$ channel with all expected SM and SUSY backgrounds at a 
favourable point in parameter space $m_A$ = 350~GeV and $tan\beta$= 5 and with the following 
MSSM parameters:  $M_1$ = 60~GeV, $M_2$ = 120~GeV, $\mu$ = -500~GeV, $M_{\tilde{q},\tilde{g}}$ = 1000~GeV, $M_{\tilde{\ell}}$ = 250~GeV and $A_t$ = 0.
The expected $m_A$-$tan\beta$ reach for 30~$fb^{-1}$ and 100~$fb^{-1}$ is shown 
in Fig. \ref{fig:chi2chi2}. 
This parameter choice does not affect 
the discovery ranges for the Higgs decays to SM particles at intermediate
and high $tan\beta$ values 
as the branching ratios do not change significantly provided 
$m_{\tilde{q}}$ and $m_{\tilde{g}}$
remain heavy \cite{spira2}.  
The expected $m_A$-$tan\beta$ reach for the $\chi^{0}_{2}\chi^{0}_{2}$ 
channel is shown in Fig. \ref{fig:chi2chi2}
for 30~$fb^{-1}$ and 100~$fb^{-1}$. 
The figure also illustrates that if the neutralinos are heavier ($M_2$ =
180 GeV) the reach is 
reduced because of the mass threshold effect.\\
Figure \ref{fig:maxmix_100fb_susy} shows the expected $m_A$-$tan\beta$
reach for the heavy SUSY Higgs bosons including this channel and the
channels with
SM particle decays for 100~$fb^{-1}$ assuming $M_1$ = 90~GeV, $M_2$ =
180~GeV, $\mu$ = 500~GeV, $M_{\tilde{q},\tilde{g}}$ = 1000~GeV and
$M_{\tilde{\ell}}$ = 250~GeV. 
The complementarity of searches of $A$ and $H$ decays
to SM and SUSY particles is clear comparing Figs. \ref{fig:maxmix_100fb}
and \ref{fig:maxmix_100fb_susy}.

\vspace{ 3mm}
\section{Conclusions}

The expected discovery ranges for the MSSM Higgs bosons are summarized in 
Figs. \ref{fig:nomix_30fb} - \ref{fig:maxmix_100fb_susy}.
Only the main discovery channels investigated in detail in CMS are shown and 
discussed in this note. The whole parameter space not excluded by LEP 
is expected to be covered with the light Higgs decay modes $h \ra \gamma\gamma$ and 
$h \ra b\overline{b}$ in the no-mixing case already with 30~$fb^{-1}$. If the stop mixing is maximal 
a tiny corner of the parameter space between the discovery ranges for the light charged 
Higgs in $t\overline{t}$ events and for the $h \ra b\overline{b}$ channel
around $m_A \sim$ 120~GeV and $tan\beta \sim$ 5 
is still left unexplored with 30~$fb^{-1}$. With 60~$fb^{-1}$ the $h \ra b\overline{b}$ 
channel extends towards lower $m_A$ values and closes this area. The heavy neutral MSSM Higges are
expected to be discovered at high $tan\beta$  in the $H, A \ra \mu\mu$ and $H, A \ra \tau\tau$
decays. The final states of 2 $leptons$, $lepton + \tau~jet$ and 2 $\tau~jets$
 are investigated for
$H, A \ra \tau\tau$ decays. The 2 $\tau~jet$ final states are found to 
extend the reach 
significantly towards the high Higgs masses and to provide the best $\tau\tau$ 
mass resolution ($\sim$ 14\%). The $A, H \ra \chi^{0}_{2}\chi^{0}_{2}  \ra 4\ell^{\pm}+X$ channel can complement the $\tau\tau$ channel at low $tan\beta$ 
provided neutralinos and sleptons are light enough. 
For the search of the charged Higgs, the
$H^{\pm} \ra \tau\nu$ decay in $gb \ra tH^{\pm}$ 
events with fully hadronic final
states is found to be the most favourable.

Investigations on several other channels, not discussed above, 
are in progress in CMS. Among the most interesting ones are 
the studies on the weak boson fusion channels
$qq \ra qqH$  which are essential for the measurement of the $WWH$ 
and $\tau\tau H$ couplings and the total width of the 
Higgs boson \cite{zeppenfeld}. These channels with $h \ra \tau\tau$ and 
$H \ra \tau\tau$ decays could possibly cover the whole MSSM parameter space \cite{zeppenfeld2}. Promising results are also
obtained for the $H \ra \gamma\gamma$ channel with a fast
simulation method \cite{dubinin}. As the event rates in all these $qqH$ 
channels are small, and as the 
Higgs mass reconstruction and the forward jet tagging are highly detector sensitive, 
a full simulation is really needed before definite conclusions can be made.
 Such studies are now under way 
for the $H(h) \ra \tau\tau$ \cite{sasha2},  
$H(h) \ra WW \ra \ell\ell\nu\nu$ \cite{sasha3} and $h \ra 
\chi^{0}\chi^{0}$ (invisible Higgs) \cite{mazumdar} channels.

\vspace{ 3mm}
\section{Acknowledgements}
We would like to thank Salavat Abdullin for the help in various 
simulation aspects.

\newpage

\clearpage



\clearpage

\begin{2figures}{t}
  \resizebox{\linewidth}{80 mm}{\includegraphics{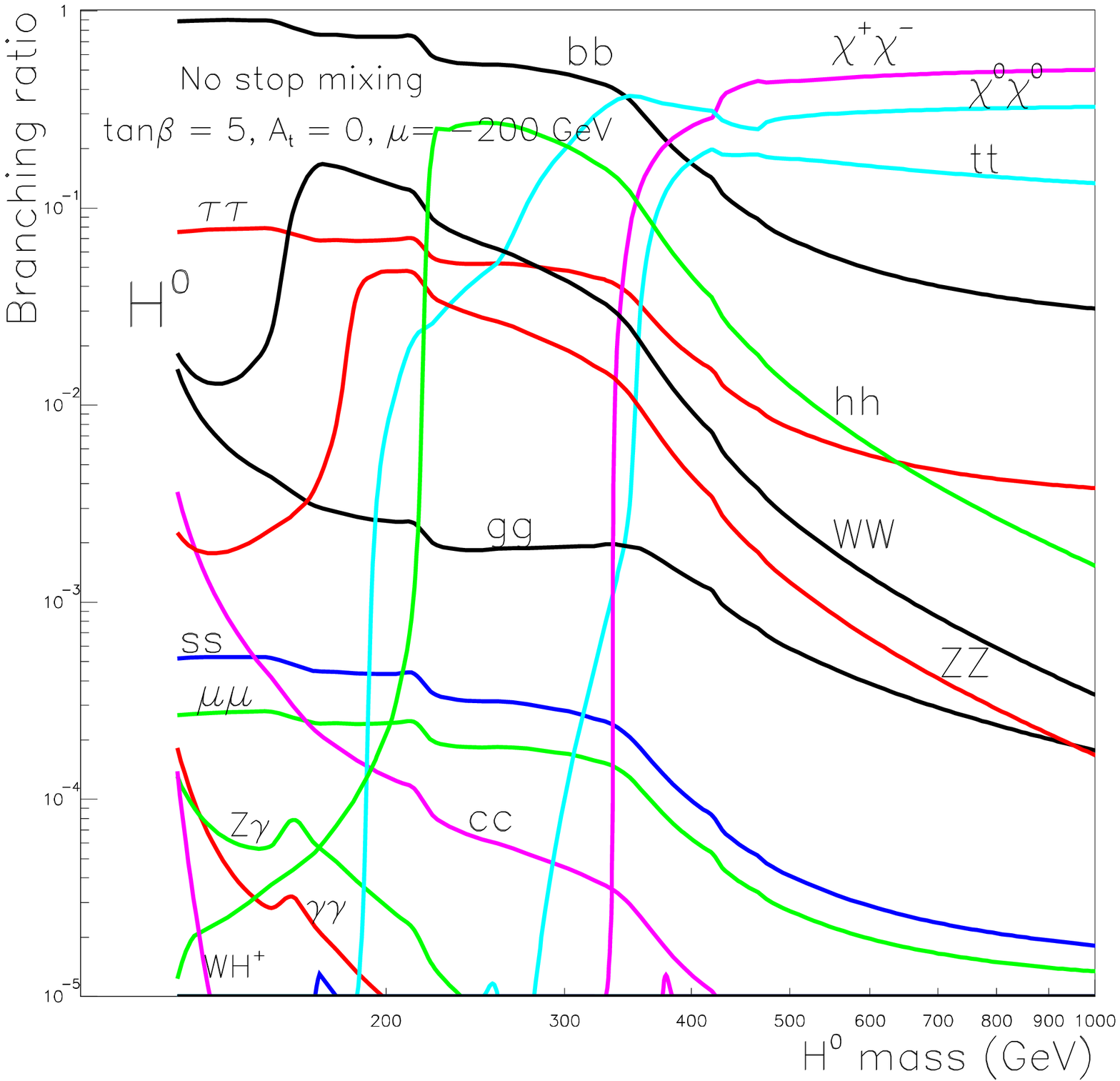}} &
  \resizebox{\linewidth}{80 mm}{\includegraphics{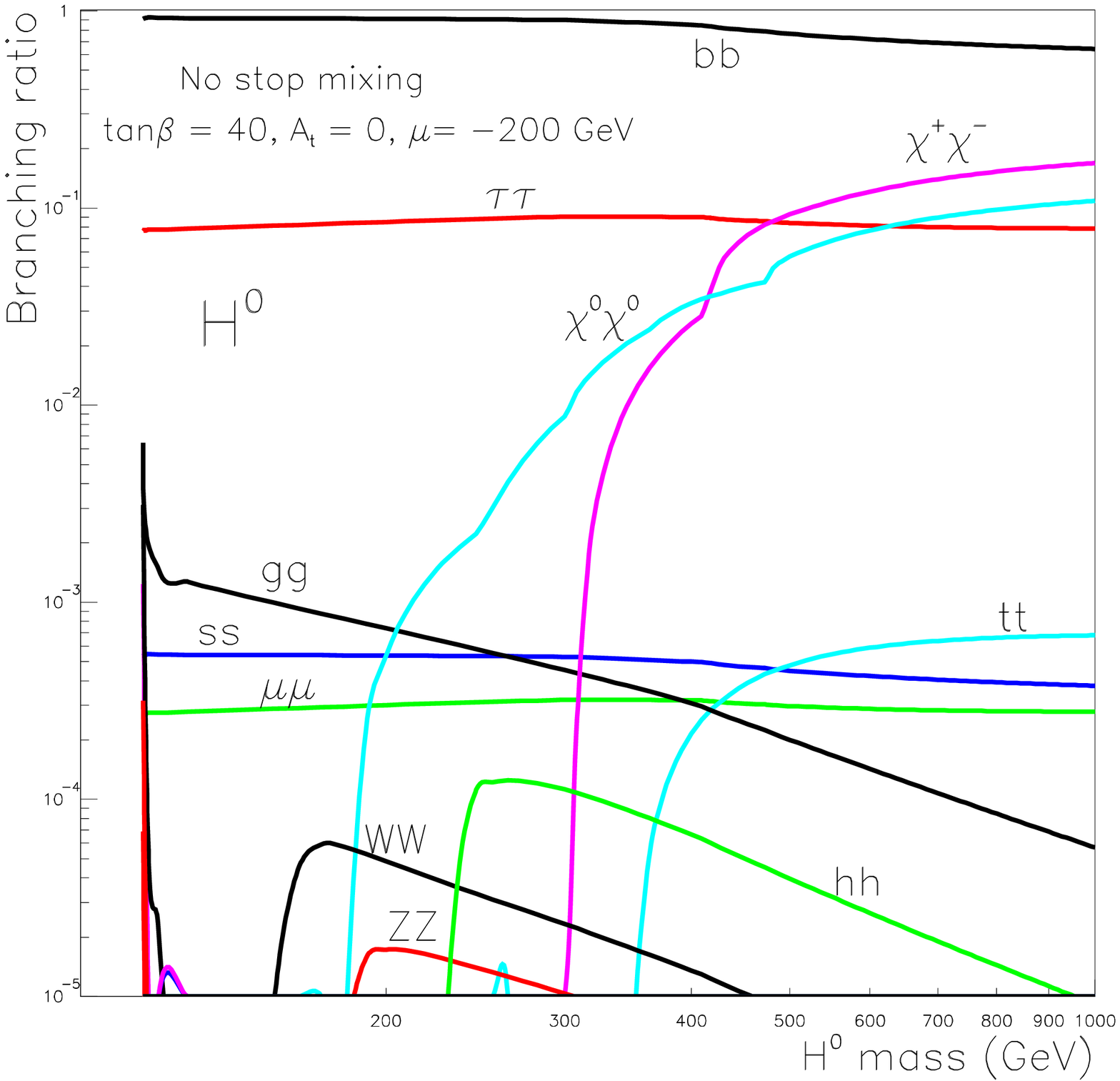}} \\
  \caption{Branching ratio for $H^0$ as a function of $m_{H}$
  for $tan\beta$ = 5 calculated with HDECAY \cite{hdecay}. 
  The SUSY parameters are taken to
  be $A_t$ = 0 (no-mixing), $M_2$ = 200~GeV, $\mu$ = -200~GeV and  
  $M_{\tilde{q},\tilde{\ell},\tilde{g}}$ = 1 TeV.} 
  \label{fig:br_hh_tanb5} &
  \caption{Branching ratio for $H^0$ as a function of $m_{H}$
           for $tan\beta$ = 40 calculated with HDECAY \cite{hdecay}. 
           The SUSY parameters are taken to
           be $A_t$ = 0 (no-mixing), $M_2$ = 200~GeV, $\mu$ = -200~GeV and  
           $M_{\tilde{q},\tilde{\ell},\tilde{g}}$ = 1 TeV.}
  \label{fig:br_hh_tanb40} \\
\end{2figures}

\begin{2figures}{b}
  \resizebox{\linewidth}{80 mm}{\includegraphics{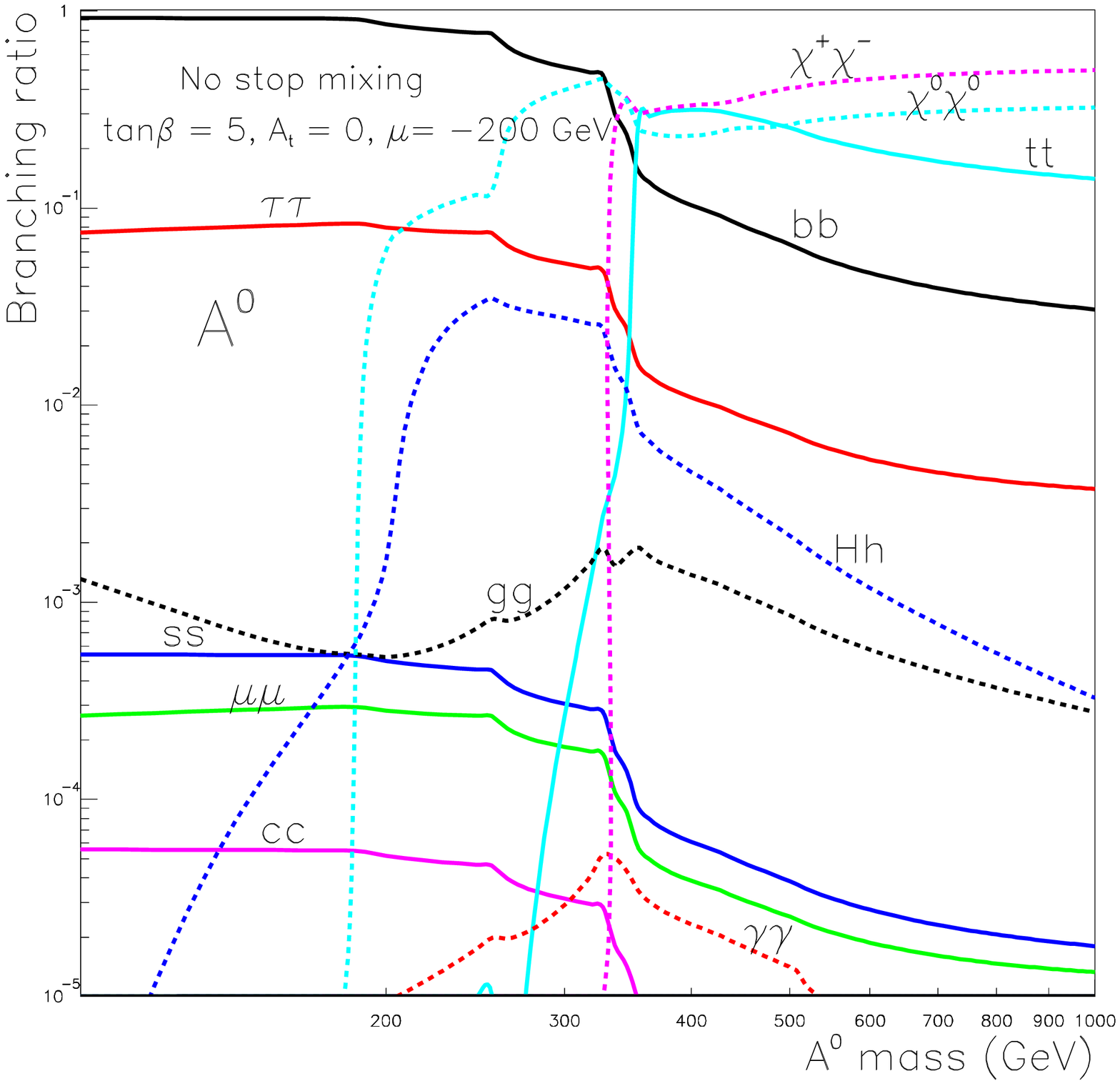}} &
  \resizebox{\linewidth}{80 mm}{\includegraphics{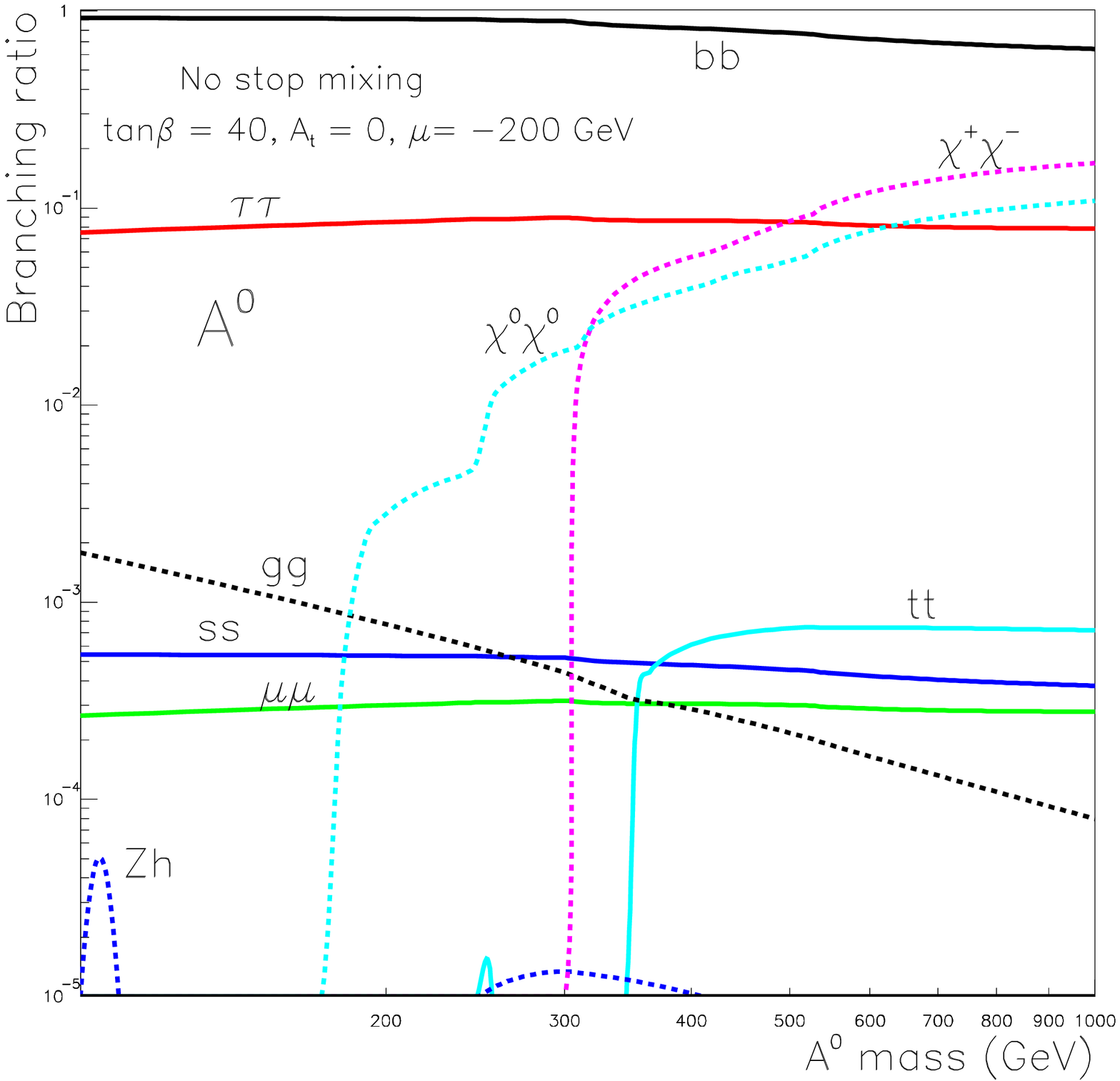}} \\
  \caption{Branching ratio for $A^{0}$ as a function of $m_{A}$
           for $tan\beta$ = 5 calculated with HDECAY \cite{hdecay}. 
           The SUSY parameters are taken to
           be $A_t$ = 0 (no-mixing), $M_2$ = 200~GeV, $\mu$ = -200~GeV and  
           $M_{\tilde{q},\tilde{\ell},\tilde{g}}$ = 1 TeV.} 
  \label{fig:br_a0_tanb5} &
  \caption{Branching ratio for $A^{0}$ as a function of $m_{A}$
           for $tan\beta$ = 40 calculated with HDECAY \cite{hdecay}. 
           The SUSY parameters are taken to
           be $A_t$ = 0 (no-mixing), $M_2$ = 200~GeV, $\mu$ = -200~GeV and  
           $M_{\tilde{q},\tilde{\ell},\tilde{g}}$ = 1 TeV.}
  \label{fig:br_a0_tanb40} \\
\end{2figures}

\clearpage
%

\begin{2figures}{t}
  \resizebox{\linewidth}{80 mm}{\includegraphics{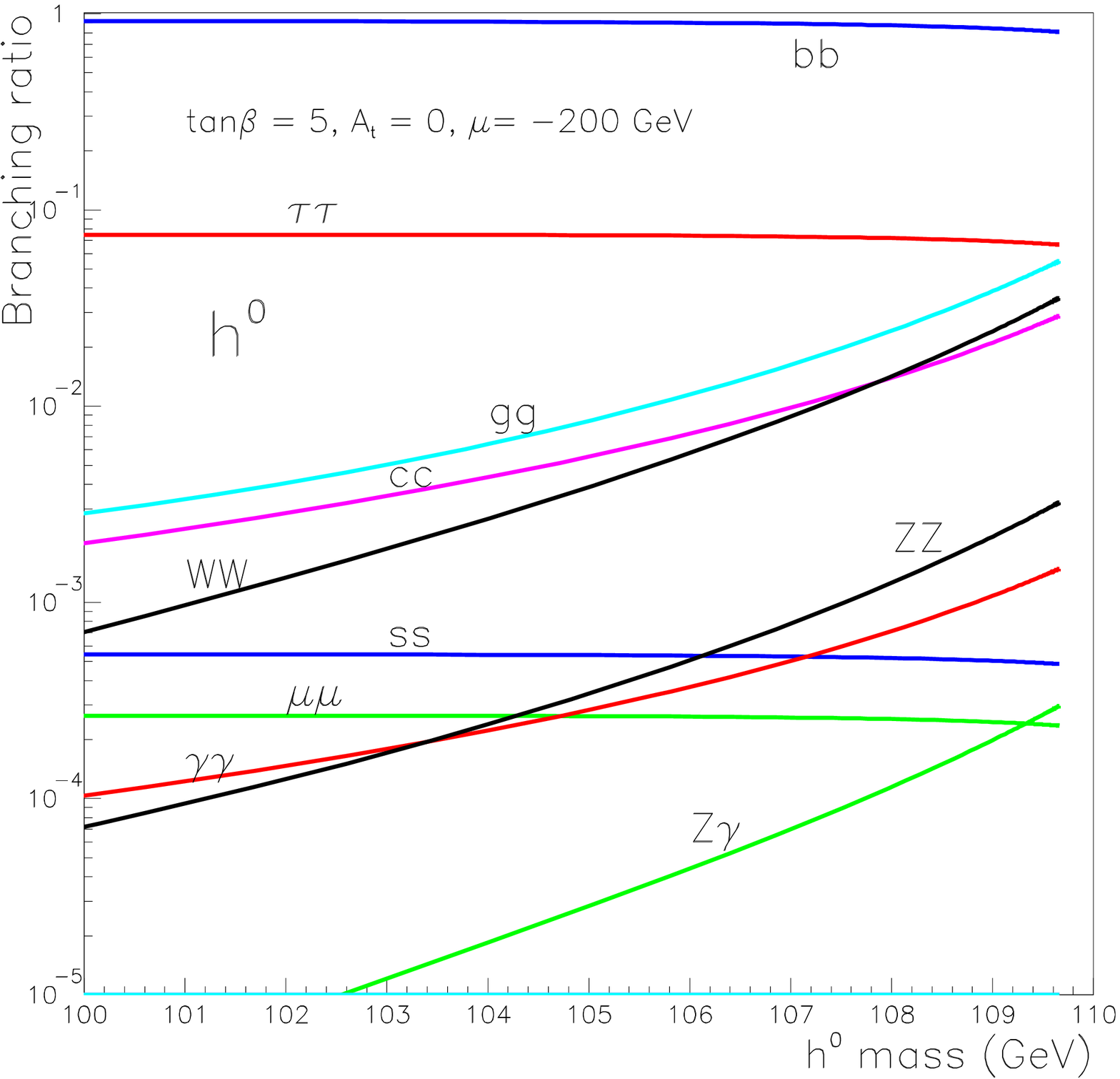}} &
  \resizebox{\linewidth}{80 mm}{\includegraphics{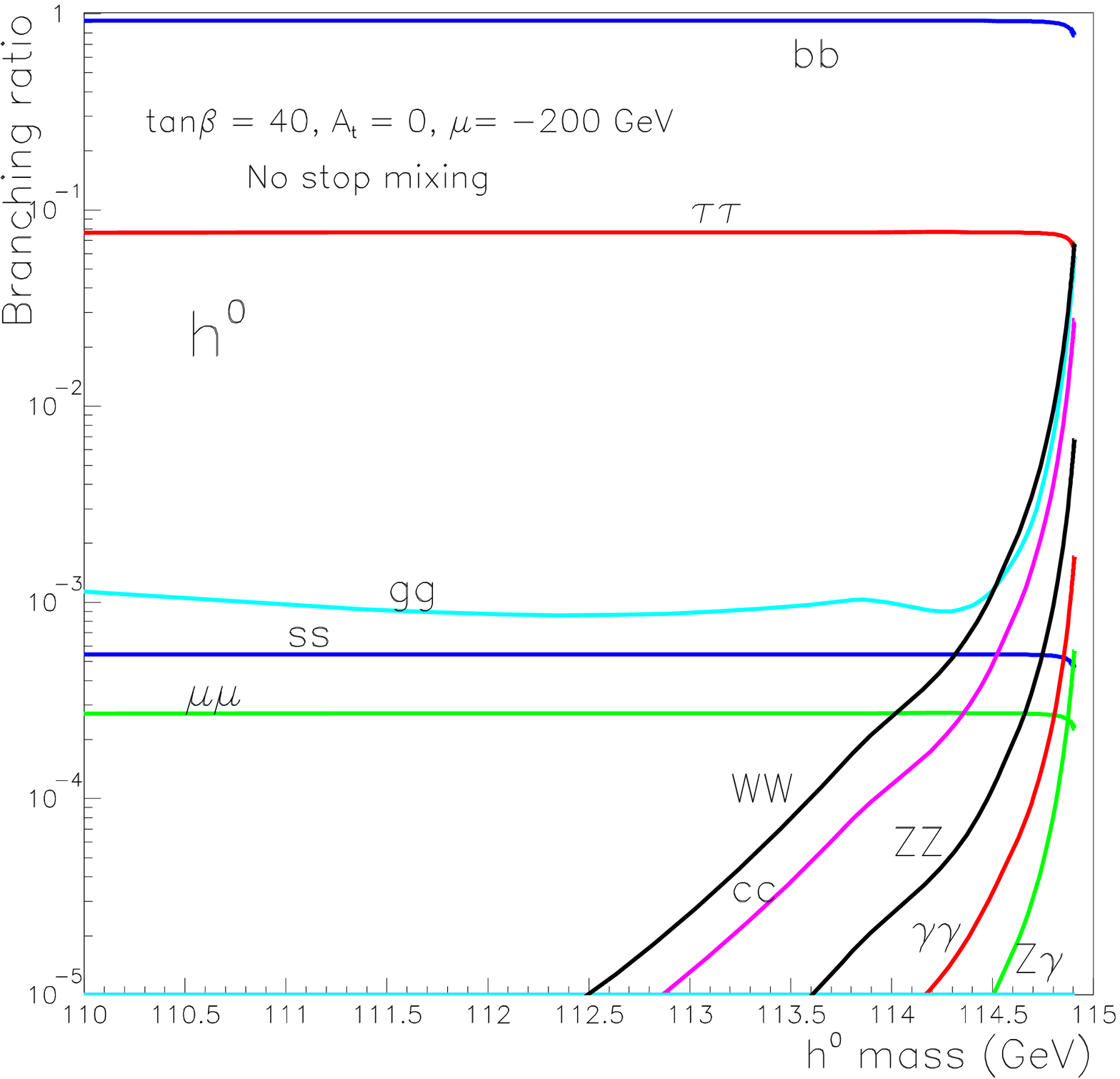}} \\
  \caption{Branching ratio for $h^0$ as a function of $m_{h}$
  for $tan\beta$ = 5 calculated with HDECAY \cite{hdecay}. 
  The SUSY parameters are taken to
  be $A_t$ = 0 (no-mixing), $M_2$ = 200~GeV, $\mu$ = -200~GeV and  
  $M_{\tilde{q},\tilde{\ell},\tilde{g}}$ = 1 TeV.} 
  \label{fig:br_hl_tanb5} &
  \caption{Branching ratio for $h^0$ as a function of $m_{h}$
           for $tan\beta$ = 40 calculated with HDECAY \cite{hdecay}. 
           The SUSY parameters are taken to
           be $A_t$ = 0 (no-mixing), $M_2$ = 200~GeV, $\mu$ = -200~GeV and  
           $M_{\tilde{q},\tilde{\ell},\tilde{g}}$ = 1 TeV.}
  \label{fig:br_hl_tanb40} \\
\end{2figures}

\begin{2figures}{b}
  \resizebox{\linewidth}{80 mm}{\includegraphics{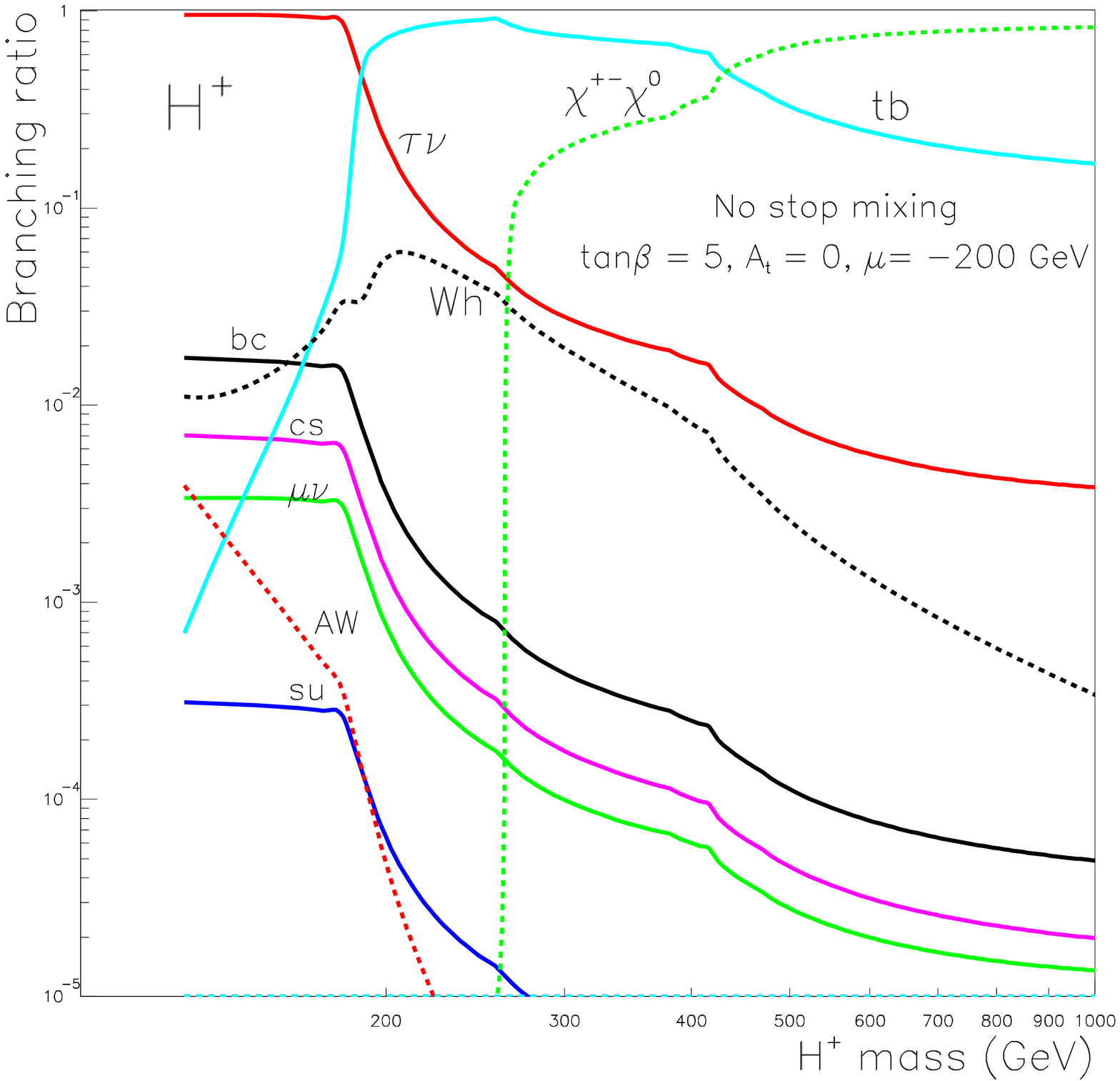}} &
  \resizebox{\linewidth}{80 mm}{\includegraphics{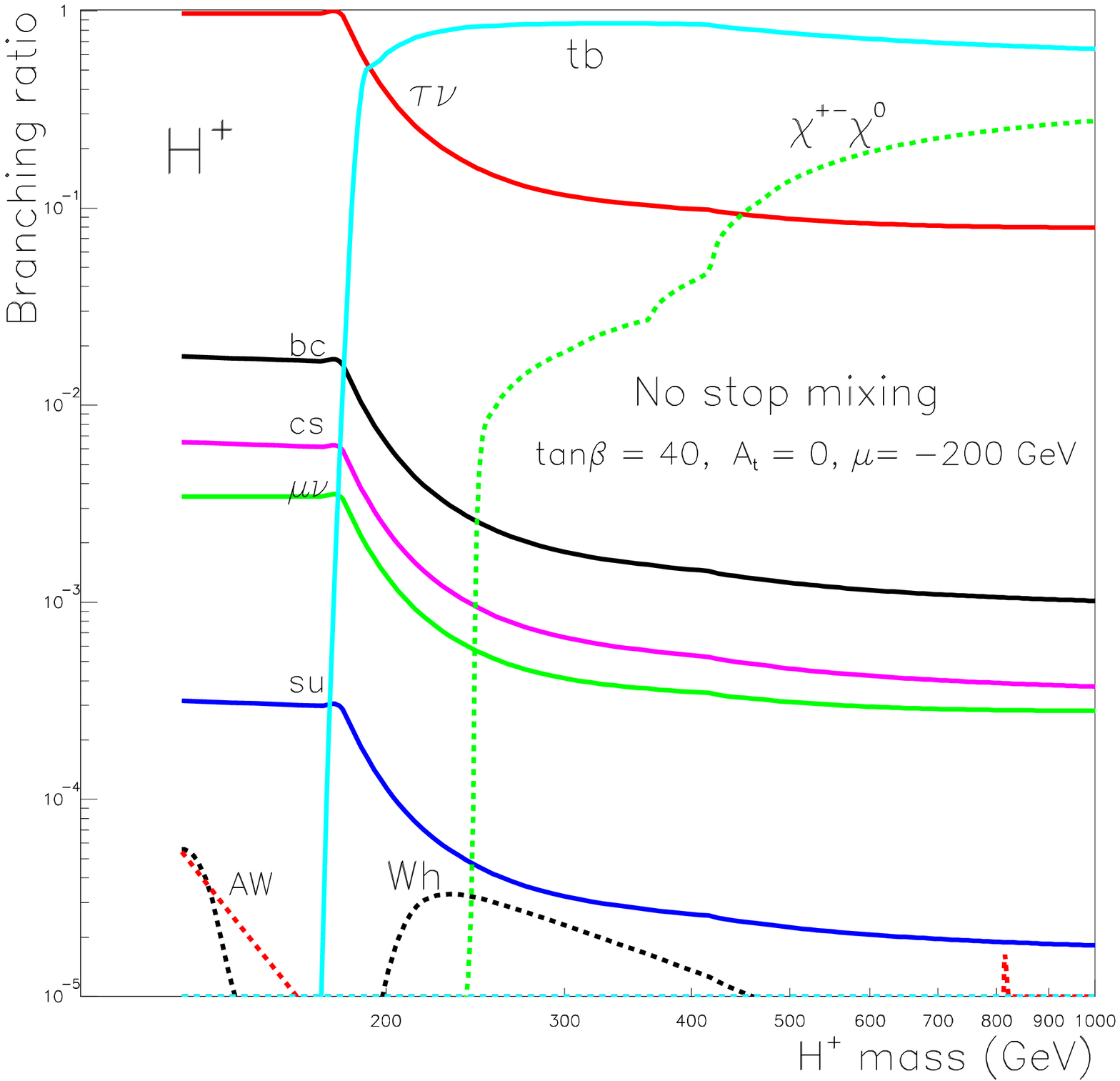}} \\
  \caption{Branching ratio for $H^{\pm}$ as a function of $m_{H^{\pm}}$
           for $tan\beta$ = 5 calculated with HDECAY \cite{hdecay}. 
           The SUSY parameters are taken to
           be $A_t$ = 0 (no-mixing), $M_2$ = 200~GeV, $\mu$ = -200~GeV and  
           $M_{\tilde{q},\tilde{\ell},\tilde{g}}$ = 1 TeV.} 
  \label{fig:br_hplus_tanb5} &
  \caption{Branching ratio for $H^{\pm}$ as a function of $m_{H^{\pm}}$
           for $tan\beta$ = 40 calculated with HDECAY \cite{hdecay}. 
           The SUSY parameters are taken to
           be $A_t$ = 0 (no-mixing), $M_2$ = 200~GeV, $\mu$ = -200~GeV and  
           $M_{\tilde{q},\tilde{\ell},\tilde{g}}$ = 1 TeV.}
  \label{fig:br_hplus_tanb40} \\
\end{2figures}

\clearpage
%
%

\clearpage

\begin{2figures}{b}
  \resizebox{\linewidth}{80 mm}{\includegraphics{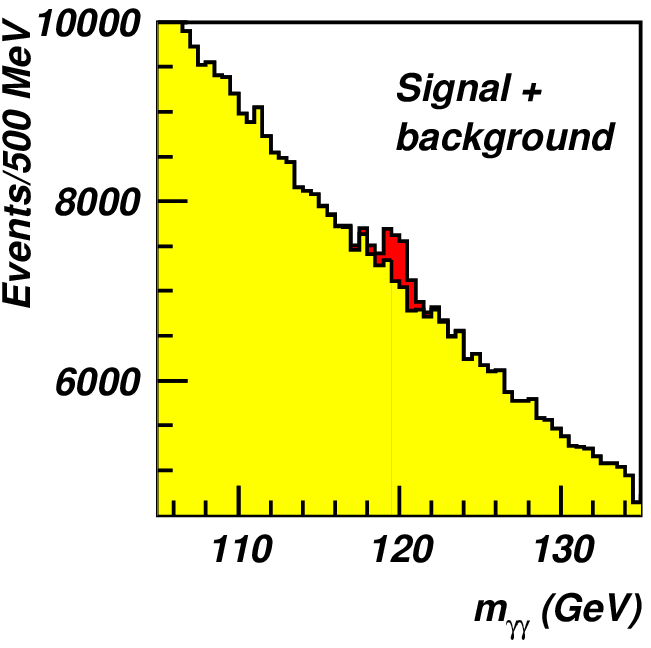}} &
  \resizebox{\linewidth}{70 mm}{\includegraphics{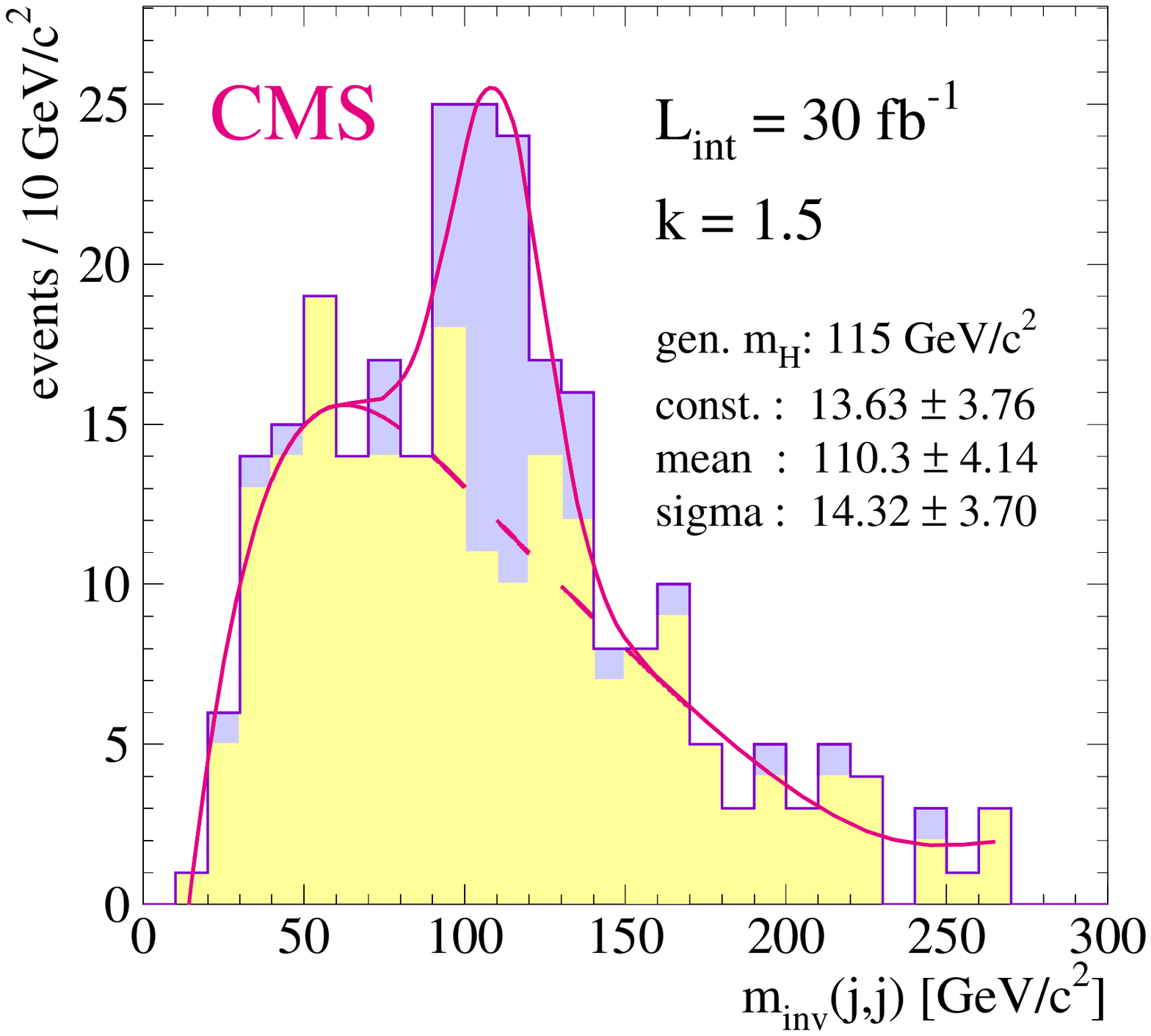}} \\
  \caption{Reconstructed Higgs mass for $H \ra \gamma\gamma$ 
           superimposed on the total background for $m_H$ = 120~GeV
           with 100~$fb^{-1}$.} 
  \label{fig:hgamma} &
  \caption{Reconstructed Higgs mass for $t\overline{t}H$,
           $H \ra b\overline{b}$ 
           superimposed on the total background for $m_H$ = 115~GeV
           with 30~$fb^{-1}$.}
  \label{fig:hbb} \\
\end{2figures}

\clearpage

\clearpage

\clearpage

\clearpage

\begin{2figures}{t}
  \resizebox{\linewidth}{80 mm}{\includegraphics{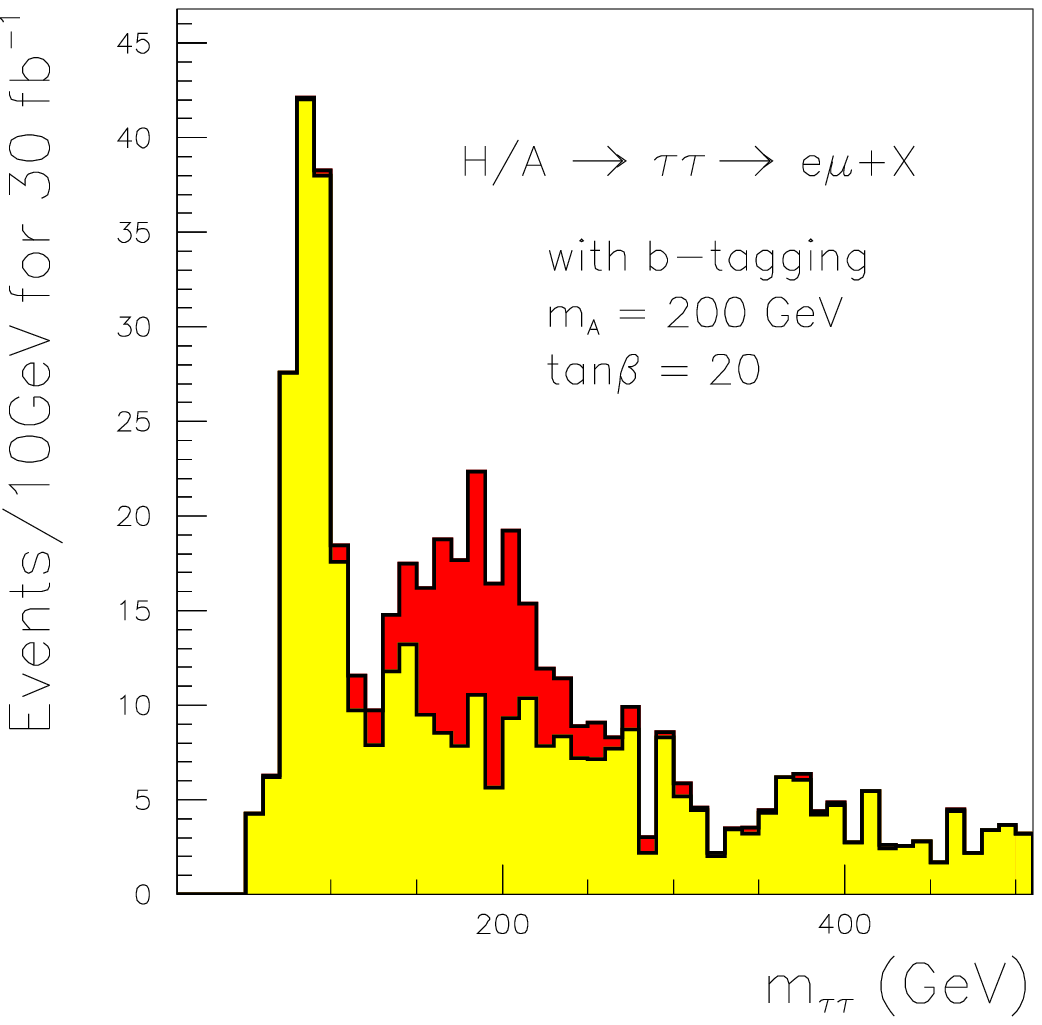}} &
  \resizebox{\linewidth}{80 mm}{\includegraphics{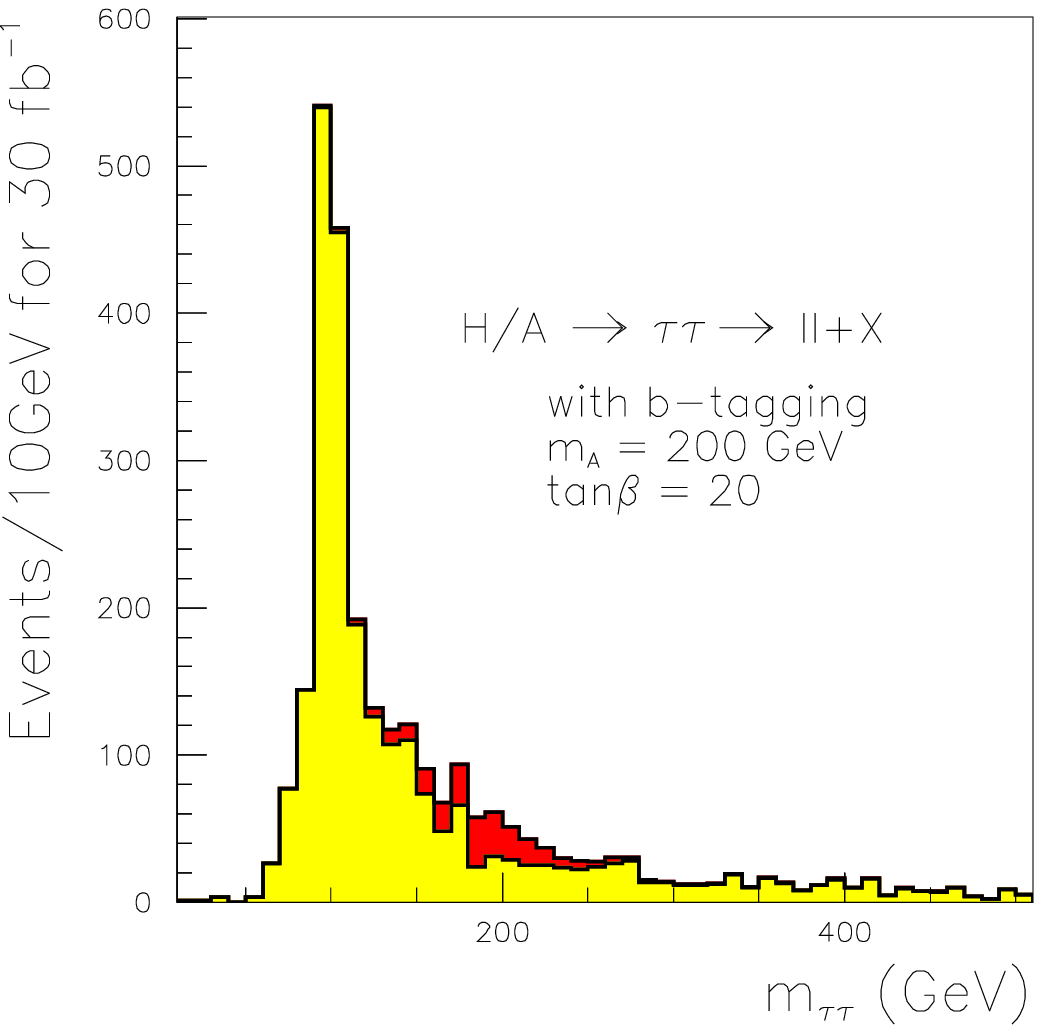}} \\
  \caption{Reconstructed Higgs mass for $H,A \ra \tau\tau \ra e\mu$ 
           superimposed on the total background for $m_A$ = 200~GeV
           and $tan\beta$ = 20 for 30~$fb^{-1}$ \cite{sami}.} 
  \label{fig:hmass_emu} &
  \caption{Reconstructed Higgs mass for $H,A \ra \tau\tau \ra \ell^+\ell-$
           superimposed on the total background for $m_A$ = 200~GeV
           and $tan\beta$ = 20 for 30~$fb^{-1}$ \cite{sami}.}
  \label{fig:hmass_ll} \\
\end{2figures}

\begin{2figures}{b}
  \resizebox{\linewidth}{80 mm}{\includegraphics{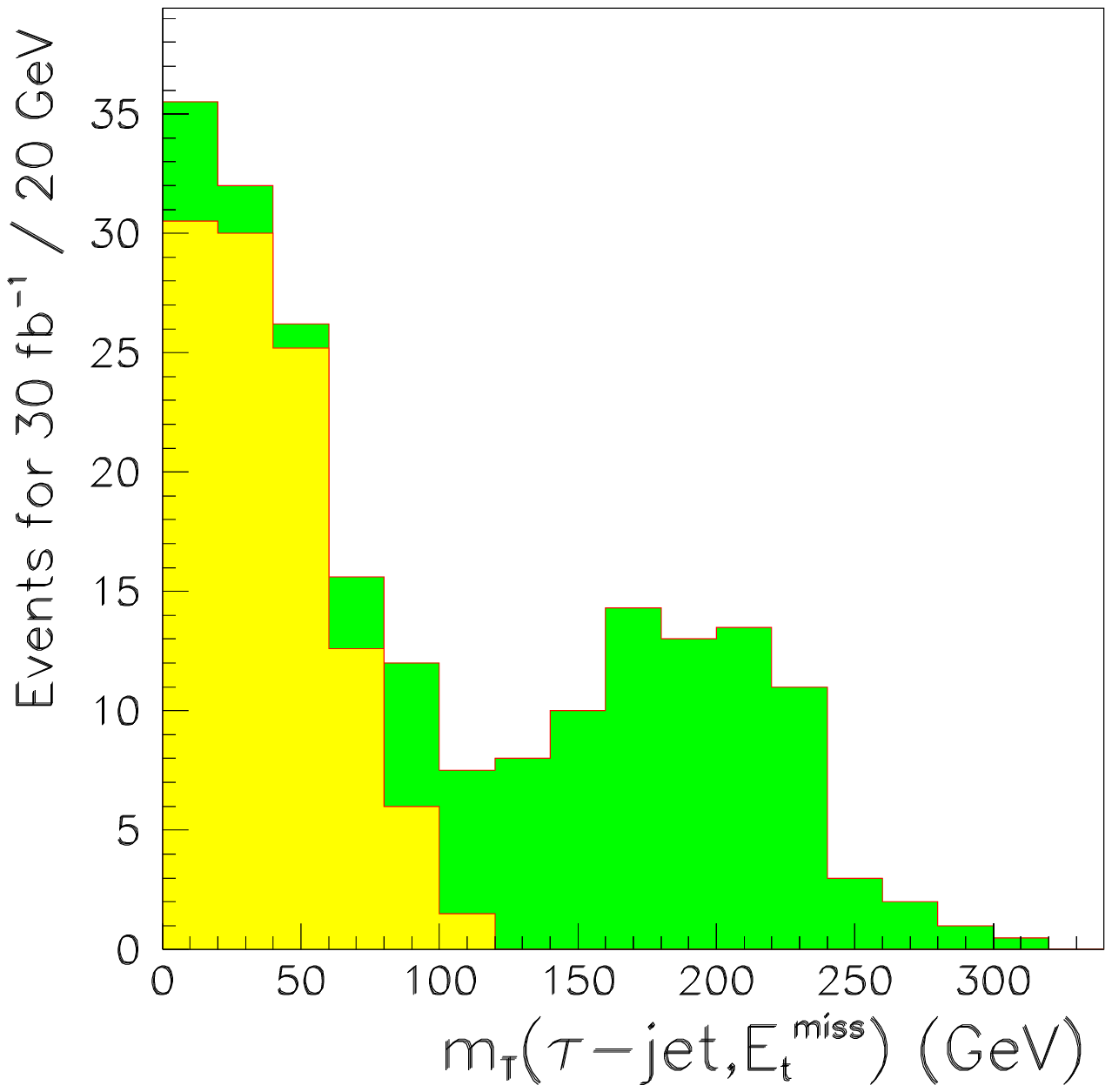}} &
  \resizebox{\linewidth}{80 mm}{\includegraphics{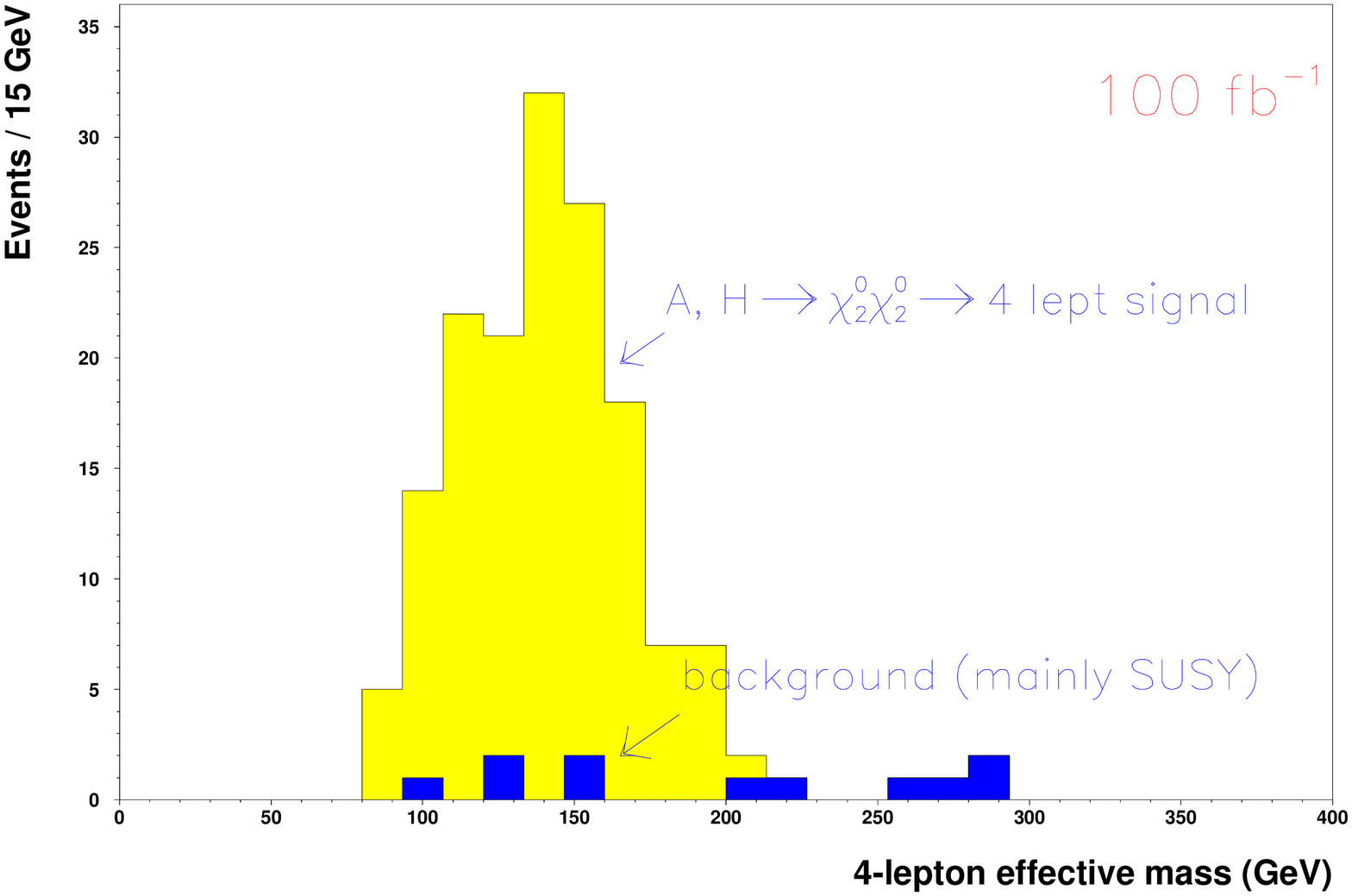}} \\
  \caption{Reconstructed transverse mass for the charged Higgs in
           $tH^{\pm}$, $H^{\pm} \ra \tau\nu$ with hadronic $\tau$ and 
           top decays  
           superimposed on the total background for $m_{H^{\pm}}$ = 217~GeV
           and $tan\beta$ = 40 for 30~$fb^{-1}$} 
  \label{fig:hplus_mass} &
  \caption{Invariant 4$\ell^{\pm}$ mass for $A, H \ra \chi^{0}_{2}\chi^{0}_{2} 
           \ra 4\ell^{\pm}+X$  superimposed on the total SM and SUSY backgrounds
           for $m_A$ = 350~GeV and $tan\beta$= 5 and with $M_1$ = 60~GeV, 
          $M_2$ = 120~GeV, $\mu$ = -500~GeV, $M_{\tilde{q},
           \tilde{g}}$ = 1000~GeV, $M_{\tilde{\ell}}$ = 250~GeV and $A_t$ = 0}
  \label{fig:filip_mass} \\
\end{2figures}

\clearpage

\newpage
\begin{thebibliography}{15}
\vspace{3mm}



\bibitem {LEPII} The ALEPH, DELPHI, L3 and OPAL Collaborations, and the LEP
Higgs Working Group, CERN-EP/2001-055 and hep-ex/0107030.

\bibitem {junk} T. Junk and V. Ruhlmann-Kleider, Private communication.

\bibitem {LEPE} The LEP Electroweak Working Group, A Combination 
of Preliminary Electroweak Measuremets and Constraints in the Standard Model, CERN-EP/2001- in preparation, presented by D. Charlton, EPS HEP 2001, 
Budapest, Hungary, July 12-18, 2001.

\bibitem {puljak} Ivica Puljak, Potentiel de decouverte du boson de Higgs
dans le canal $H \ra ZZ^* \ra 4\ell^{\pm}$ avec le detecteur CMS. Contribution
a la construction du calorimetre electromagnetique de CMS, These de Doctorat
de l'Universite Paris VI, Septembre 2000.
 
\bibitem {dittmar} M. Dittmar, hep-ex/9901009.

\bibitem{pythia57} T. Sjostrand, Comp.Phys.Comm. 82 (1994) 74;
CERN-TH.6488/92; CERN-TH.7112/93.

\bibitem{pythia}  T. Sjostrand, P. Eden, Ch. Friberg, L. Lonnblad, C. Miu,
S. Mrenna and E. Norrbin, High-Energy-Physics Event Generation with PYTHIA 6.1,
Computer Phys. Commun. 135 (2001) 238.

\bibitem{gunion} J.F. Gunion, A. Stange and S. Willenbrock,
Weakly-coupled Higgs Bosons, UCD-95-28,hep-ph/9602238

\bibitem{tauola} S. Jadach, Z. Was, R. Decker, M. Jezabek and J.H. Kuhn, 
 CERN-TH-6793, 1992.  

\bibitem{hdecay} A. Djouadi, J. Kalinowski and M. Spira, HDECAY: a program
for Higgs Boson Decays in the Standard Model and its Supersymmetric 
expension, hep-ph/9704448.

\bibitem{cmsjet} S. Abdullin, A. Khanov and N. Stepanov, CMSJET, CMS
TN/94-180.

\bibitem {LEP} M. Carena, S. Heynemeyer, C.E.M. Wagner and G. Weiglein, 
Suggestions for improved benchmark scenarios for Higgs-boson searches at
LEP2, CERN-TH/99-374, DESY 99-186, hep-ph/9909435.

\bibitem {mupar} Muon (g-2) Collaboration, H. N. Brown et al., Phys. Rev. Lett.
86 (2001) 2227.

\bibitem {tdr:tracker} 
CMS Collaboration, The tracker project, Technical Design Report,
 CERN/LHCC 98-6, CMS TDR 5, 26 February 1998.

\bibitem {higlu} M. Spira, Higlu : a program for the calculation of the 
total Higgs production cross section at hadron colliders
 via gluon fusion including QCD corrections, hep-ph/9510347.

\bibitem{pphtt} M. Spira, PPHTT, program to calculate cross sections 
for processes $gg \ra b\overline{b}H_{SUSY}, t\overline{t}H_{SUSY}$, http://m.home.cern.ch/m/mspira/www/hqq/.

\bibitem {tdr:ecal} 
CMS Collaboration, The electromagnetic calorimeter project, Technical Design Report,
 CERN/LHCC 97-33, CMS TDR 4, 15 December 1997.

\bibitem{seez} C. Seez, privite communication.

\bibitem{lassila}
Katri Lassila-Perini, Discovery Potential of the Standard Model Higgs in CMS 
at the LHC, Diss. ETH N.12961.

\bibitem{djouadi} A.Djouadi, Phys. Lett. B435 (1998) 101

\bibitem{hep-ph/9806315} A. Djouadi, Squark effects on Higgs 
boson production and decay at the LHC, hep-ph/9806315

\bibitem{loops} R. Kinnunen, S. Lehti,  A. Nikitenko and S. Rantala, Effects of large 
mixing and light stop for $h \ra \gamma\gamma$ in MSSM,  CMS NOTE 2000/043.

\bibitem {mile} M. Dzelalija, Z. Antunovic, D. Denegri and R. Kinnunen,
Study of the Associated production Modes $Wh, t\overline{t}$ in the Minimal
Supersymmetric Standard Model in CMS, CMS TN/96-091.

\bibitem{volker}
Volker Drollinger, Reconstruction and Analysis Medhods for Searches of
Higgs Bosons in the Decay Mode $H \ra b\overline{b}$ at Hadron Colliders,
Dissertation, Institut fur Experimentelle Kernphysik, University of Karlsruhe,
Germany, July 2001.

\bibitem{volker2}
V. Drollinger, Th. Muller and D. Denegri, Searching for Higgs Bosons 
in Association with Top Quark Pairs in the $H \ra b\overline{b}$ Decay Mode,
CMS NOTE 2001/054, hep-ph/0111312.

\bibitem{dangreen}
D. Green, K. Maeshima, R. Vidal, W. Wu and S. Kunori,
A Study of ttbar + Higgs at CMS, CMS NOTE-2001/039.

\bibitem{CompHep} A. Pukhov, E. Boos, M. Dubinin, V. Edneral, V. Ilyin, D.
Kovalenko, A. Kryukov, V. Savrin, S. Shichanin and A. Semenov, CompHEP -
a package for evaluation of Feynman diagrams and integration over
multi-particle phase space, INP-MSU 98-41/542.

\bibitem{CPinterf} A. S. Belyaev, E. E. Boos, A. N. Vologdin, M. N.
Dubinin, V. A. Ilyin, A. P. Kryukov, A. E. Pukhov, A. N. Skachkova, V. I.
Savrin, A. V. Sherstnev, S. A. Shichanin, CompHEP-PYTHIA interface:
integrated package for the collision events generation based on exact
matrix elements, hep-ph/0101232

\bibitem {tdr:muon} 
CMS Collaboration, The muon project, Technical Design Report, CERN/LHCC 97-32, 
CMS TDR 3, 15 December, 1997.

\bibitem{furic} I.K. Furic and R. Kinnunen, Study of $H, A \ra \mu\mu$ in CMS, 
CMS NOTE 1998/039.

\bibitem{sami}
Sami Lehti, Prospects for the Detection of the MSSM Higgs Bosons Decaying into Tau Leptons
in the CMS Detector, Dissertation, University of Helsinki, Report Series in 
Physics, HU-P-D93, 2001.

\bibitem{h2jet} R. Kinnunen and D. Denegri, Study of $H, A \ra \tau\tau 
\ra h^+ + h- + X$ in CMS,  CMS NOTE 1999/037.

\bibitem{trigger1} S. Eno, W. Smith, S. Dasu, R. Kinnunen and A. Nikitenko, A Study
of a First and Second Level Tau Trigger, CMS NOTE 2000/055.

\bibitem{trigger3} D. Kotlinski, A. Nikitenko and  R. Kinnunen, Study of a Level-3 
Tau Trigger with the Pixel Detector, CMS NOTE 2001/017.

\bibitem{sasha4} A. Nikitenko, CMS NOTE in preparation.

\bibitem{tau_vertex} A. Nikitenko and L. Wendland, Talk in the btau Meeting, 21 August, 2001.

\bibitem{sasha} A. Nikitenko, S. Kunori and R. Kinnunen, 
Missing Transverse Energy Measurement with Jet Energy Corrections, CMS NOTE 2001/040.

\bibitem{hljet} R. Kinnunen and A. Nikitenko, Study of $H, A \ra \tau\tau \ra 
\ell + \tau jet + E_t^{miss}$ in CMS,  CMS NOTE 1997/106. 

\bibitem{dproy} D.P. Roy, The Hadronic Tau Decay Signature of Heavy Charged Higgs 
Boson at LHC, Phys. Lett. B459(1999)607.

\bibitem{hplus} R. Kinnunen, Study of Heavy Charged Higgs in $pp \ra tH^{\pm}$ with
$H^{\pm} \ra \tau\nu$ in CMS, CMS NOTE 2000/045.

\bibitem{manas} S. Banerjee, M. Mainty, Search for Charged Higgs in Top Decays
in CMS, CMS NOTE 2000/039.

\bibitem{moretti} S. Moretti and D.P. Roy, Detecting Heavy Charged Higgs 
Boson at LHC with Triple B-tagging, hep-ph/9909435.

\bibitem{serguei} S. Slabospitsky, CMS NOTE in preparation. 

\bibitem{filip} S. Abdullin, D. Denegri and F. Moortgat, Observability of MSSM 
Higgs bosons via sparticle decay modes in CMS, CMS NOTE 2001/042;   \\
F. Moortgat, Observability of MSSM Higgs bosons decaying to sparticles at the LHC, 
CMS CR 2001/005.


\bibitem{nikita} P. Salmi, R. Kinnunen  and N. Stepanov, 
CMS NOTE in preparation.

\bibitem{spira2} M. Spira, private communication.

\bibitem{zeppenfeld} D. Zeppenfeld, R. Kinnunen, A. Nikitenko and E. Ricter-Was,
Phys. Rev. D 62 (2000) 013009.

\bibitem{zeppenfeld2} D. Rainwater and D. Zeppenfeld, Journal of High Energy Phys. 
12 (1997) 005.

\bibitem{dubinin} M. Dubinin, Higgs Boson Signal in the Reaction
 $pp \ra  Gamma~Gamma + 2~Forward~Jets$  , CMS NOTE 2001/022. 

\bibitem{sasha2} A. Nikitenko, Tau Trigger for Higgs Studies at CMS, Invited Talk
given in Higgs and Supersymmetry, Orsay, March 19-22, 2001.

\bibitem{sasha3} A. Nikitenko et al., CMS NOTE in preparation.

\bibitem{mazumdar} K. Mazumdar and A. Nikitenko, CMS NOTE in preparation.  

\end{thebibliography}
\end{document}